\newcommand{\partdev}[2]{\frac{\partial #1}{\partial #2}}
\newcommand{\diffix}[3]{\left(\frac{\partial #1}{\partial #2}\right)_{#3}}
\newcommand{\dd}{\mathrm{d}}

 \documentclass[%
  aip,jcp,
 amsmath,amssymb,
  reprint,%
 ]{revtex4-1}

\usepackage{graphicx}% Include figure files
\usepackage{dcolumn}% Align table columns on decimal point
\usepackage{bm}% bold math
\usepackage{mathtools}% For using dcases
\usepackage{multirow}
\usepackage{makecell}

\usepackage[dvipsnames]{xcolor}

\begin{document}

\preprint{APS/123-QED}

\title{Statistical mechanics of crystal nuclei of hard spheres}% Force line breaks with \\

\author{Marjolein de Jager}
\affiliation{Soft Condensed Matter and Biophysics, Debye Institute for Nanomaterials Science, Utrecht University, Utrecht, Netherlands}
\author{Carlos Vega}
\affiliation{Departamento de Qu\'imica F\'isica, Facultad de Ciencias Qu\'imicas, Universidad Complutense de Madrid, 28040 Madrid, Spain}
\author{Pablo Montero de Hijes}
\affiliation{University of Vienna, Faculty of Physics, Boltzmanngasse 5, A-1090 Vienna, Austria}
\affiliation{University of Vienna, Faculty of Earth Sciences, Geography and Astronomy, Josef-Holaubuek-Platz 2, 1090 Vienna, Austria}
\author{Frank Smallenburg}
\affiliation{Universit\'e Paris-Saclay, CNRS, Laboratoire de Physique des Solides, 91405 Orsay, France}
\author{Laura Filion}
\affiliation{Soft Condensed Matter and Biophysics, Debye Institute for Nanomaterials Science, Utrecht University, Utrecht, Netherlands}%Lines break automatically or can be forced with \\

% \collaboration{CLEO Collaboration}%\noaffiliation

\date{\today}% It is always \today, today,

\begin{abstract}
In the study of crystal nucleation via computer simulations, hard spheres are arguably the most extensively explored model system. Nonetheless, even in this simple model system, the complex thermodynamics of crystal nuclei can sometimes give rise to counterintuitive results, such as the recent observation that the pressure inside a critical nucleus is lower than that of the surrounding fluid, seemingly clashing with the strictly positive Young--Laplace pressure we would expect in liquid droplets. Here, we re-derive many of the founding equations associated with crystal nucleation, and use the hard-sphere model to demonstrate how they give rise to this negative pressure difference.
We exploit the fact that, in the canonical ensemble, a nucleus can be in a (meta)stable equilibrium with the fluid, and measure the surface stress for both flat and curved interfaces. Additionally, we explain the effect of defects on the chemical potential inside the crystal nucleus. Lastly, we present a simple, fitted thermodynamic model to capture the properties of the nucleus, including the work required to form critical nuclei.
\end{abstract}

\maketitle

%\tableofcontents

\newcommand{\figwidthA}{0.935\linewidth}
\newcommand{\figwidthC}{0.45\linewidth}

%%%%%%%%%%%%%%%%%%%%%%%%%%%%%%%%%%%%%%%%%%%%%%%%%%%%%%%%%%%%%%%
%%%%%%%%%%%%%%%%%%%%%%%%%%%%%%%%%%%%%%%%%%%%%%%%%%%%%%%%%%%%%%%

\section{Introduction}
Hard spheres play a central role in our understanding of phase behavior, having been the focus of studies ranging from phase boundaries \cite{hoover1968melting, frenkel1984new, pusey1986phase}, to defects \cite{bennett1971studies, bowles1994cavities, pronk2001point, van2017diffusion}, to glasses \cite{speedy1998hard, pusey2009hard, zaccarelli2009crystallization, parisi2010mean}, to crystal nucleation \cite{auer2001prediction, gasser2001real, filion2010crystal, richard2018crystallizationI}.  In particular, the nucleation behavior of hard spheres has drawn significant attention (see e.g. Refs. \onlinecite{filion2011simulation, radu2014solvent, wood2018coupling, fiorucci2020effect, wohler2022hard}) due to the alarming mismatch between computationally predicted nucleation rates and experimental observations \cite{auer2001prediction, royall2023colloidal}.

Over the last few years, a number of simulation studies have taken a new route to access properties associated with crystal nucleation (see e.g. Refs. \onlinecite{statt2015finite, koss2017free, richard2018crystallizationII, gunawardana2018theoretical, montero2020interfacial, montero2020young}). Instead of focusing on the nucleation process, they have focused on equilibrium crystal nuclei in (meta)stable coexistence with their surrounding fluid.  Specifically, in the canonical ensemble, spherical crystal nuclei are stable for a range of system sizes and densities. By focusing on equilibrium nuclei, equilibrium statistical physics is guaranteed to hold, facilitating careful studies of the equilibrium structure and thermodynamic properties of crystal nuclei. A recent intriguing observation from one of these studies \cite{montero2020young} on pseudo-hard spheres showed that, counter-intuitively, the pressure inside the  crystal nucleus was lower than in the surrounding fluid. This clashes with our usual expectation of a Young--Laplace pressure which raises the pressure inside a liquid droplet with respect to the surrounding medium. Pseudo-hard spheres are not the only case where such an atypical pressure difference was seen, it was also seen in  e.g. hard spheres with short range-attractions \cite{cacciuto2004breakdown} and binary hard-sphere mixtures \cite{cacciuto2005stresses}, and good theoretical foundations exist for explaining it \cite{mullins1984thermodynamic, cacciuto2005stresses}. In particular, Mullins derived expressions for the pressure inside a crystal nucleus that is strained by its contact with the surrounding fluid \cite{mullins1984thermodynamic}. Additionally, Montero de Hijes \textit{et al.} derived a variation of the Young--Laplace equation linking the (positive) interfacial free energy to the difference in pressure between the fluid and an equilibrium bulk crystal at the same chemical potential \cite{montero2020young}. 

In this paper, we re-derive many of the founding equations associated with  crystal nucleation \cite{mullins1984thermodynamic, cacciuto2005stresses}, and apply them to one of the most fundamental model systems: monodisperse hard spheres. This paper is organized as follows: in section \ref{sec:pressure} we explore the pressure inside the crystal nucleus and show the link between the pictures of Mullins \cite{mullins1984thermodynamic} and Montero de Hijes \textit{et al.}\cite{montero2020young}, in section \ref{sec:stress} we measure the surface stress for both flat and curved interfaces, in section \ref{sec:mu} we discuss the chemical potential inside the crystal phase, in section \ref{sec:model} we propose a fitted thermodynamic model for the properties of the spherical fluid--crystal interface, and in section \ref{sec:cnt} we determine the work required to form critical nuclei.

%%%%%%%%%%%%%%%%%%%%%%%%%%%%%%%%%%%%%%%%%%%%%%%%%%%%%%%%%%%%%%%
%%%%%%%%%%%%%%%%%%%%%%%%%%%%%%%%%%%%%%%%%%%%%%%%%%%%%%%%%%%%%%%

\section{Pressure inside a crystal nucleus}
\label{sec:pressure}

An intriguing observation made in several recent papers \cite{montero2020young,montero2022thermodynamics} is that for pseudo-hard spheres, the pressure inside the crystal nucleus is found to be lower than that of the surrounding fluid. At first glance this contradicts our intuitive understanding of the pressure difference arising from the Young--Laplace equation which governs the behavior of a liquid droplet in a gas.  For such a droplet, the internal pressure is always higher than the external one with the difference proportional to the interfacial free energy.  In the case of a crystal surrounded by a fluid, however, the situation is more complex: the surface free energy is not only dependent on the amount of surface, but also on the lattice spacing of the crystal.  As such, the unexpectedly lower pressure of the crystal of pseudo-hard spheres can be attributed to the properties of the crystal--fluid interface.  

In this section, we first revisit the theory of spherical crystal nuclei \cite{mullins1984thermodynamic,cacciuto2005stresses}, and then measure the pressure inside and outside of a spherical nucleus of perfectly hard spheres. 

%%%%%%%%%%%%%%%%%%%%%%%%%%%%%%%%%%%%%%%%%%%%%%%%%%%%%%%%%%%%%%%

\subsection{Theory}
\subsubsection{Imposing thermodynamic equilibrium}
Let us start by briefly revisiting the theory of fluid--crystal phase coexistence \cite{mullins1984thermodynamic,cacciuto2005stresses}. Consider the case of a crystal nucleus inside a parent fluid phase.  We are interested in the case where the nucleus is in equilibrium with the fluid, which can be either a stable, metastable, or unstable equilibrium depending on the conditions of the system. In the canonical ensemble, it is possible for a nucleus to be in a (meta)stable equilibrium with the fluid, a feature which we will exploit in our simulations later in this work. In the grand-canonical and isobaric-isothermal (Gibbs) ensembles, the same configuration would correspond to a critical nucleus, i.e. a saddle-point in the free-energy landscape \cite{richard2018crystallizationI, rosales2020seeding}. The grand-canonical ensemble is more convenient for the theoretical treatment of systems with interfaces, and hence we will use it for the following derivation. In the Supplementary Material we include the same derivation in the canonical ensemble.

\begin{figure}[t]
    %\centering
    \includegraphics[width=0.5\linewidth]{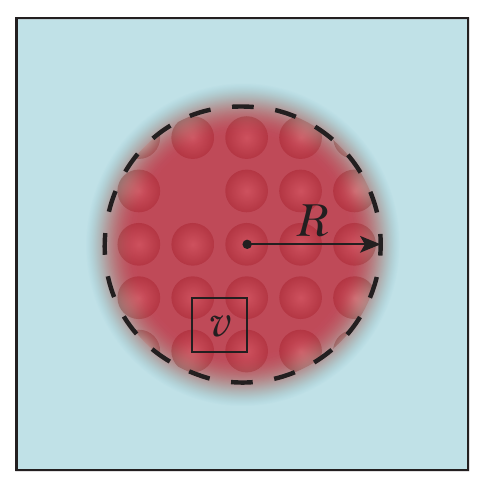}
    \caption{\label{fig:sketch} 
    Schematic representation of a spherical crystal nucleus inside a parent fluid phase. Here, $R$ indicates the radius of the dividing surface and $v$ indicates the unit cell volume of the crystal. Notice that the nucleus is depicted with a vacancy. 
    }
\end{figure}

For simplicity, we consider a spherical crystal nucleus, see Fig. \ref{fig:sketch}. In practice, the location of an interface between two coexisting phases is not unambiguously defined. Nonetheless, following Gibbs \cite{gibbs1906scientific}, it is common to define a \textit{dividing surface} between the two phases, which has zero thickness but may have a number of particles associated with it.  Using this interface, the volume of the system $V$ can be divided perfectly into the fluid volume and the crystal volume ($V = V_F + V_X$). For a spherical nucleus, the crystal volume is given by $V_X = 4\pi R^3/3$, with $R$ the radius of the nucleus for a given choice of dividing surface. The total number of particles is then given by 
\begin{equation}
    N=N_F+N_X + N_S = \rho_F V_F + \rho_X V_X + N_S, \label{eq:N}
\end{equation}
with $\rho_{F(X)}$ the number density of the fluid (crystal) phase far from the interface.
Note that, depending on the choice of dividing surface, the number of interfacial particles $N_S$ can be positive, negative, or zero. The choice of dividing surface that corresponds to $N_S=0$ is called the \textit{equimolar surface}. 

In the grand-canonical ensemble, the total grand potential $\Omega_\text{tot}$ of this system is given by
\begin{eqnarray} 
    \Omega_\text{tot}(\mu,V,T;V_X,v) &=& \Omega_F(\mu,V_F,T) + \Omega_X(\mu,V_X,T,v) \nonumber\\
    &&+ \Omega_S(\mu,R,T,v). \label{eq:omegatot}
\end{eqnarray}
Importantly, our grand potential $\Omega_\text{tot}$ depends on five variables. The first three are the thermodynamic variables defining the state of the system, namely the chemical potential $\mu$, volume $V$, and temperature $T$, which define the state point at which we examine our system. They can be regarded as external variables. Note that by necessity, $T$ and $\mu$ are homogeneous throughout the system. The last two parameters determining $\Omega_\text{tot}$ are internal variables of the system, namely the nucleus volume $V_X$ and the crystal unit cell volume $v$, which are not fixed externally. The presence of the variable $v$ accounts for the possibility of configurations where the crystal phase is strained (i.e. compressed or stretched) with respect to the equilibrium lattice spacing of a bulk crystal at chemical potential $\mu$.  For simplicity, here we assume the crystal to have cubic symmetry and only consider strains that isotropically compress or decompress the crystal, such that the unit cell retains its cubic shape. 
Note that crystals with equal $v$ but different chemical potential $\mu$ in practice correspond to crystals with the same lattice spacing, but different concentrations of vacancies and interstitials. We will return to this topic in Section \ref{sec:mu}.  Since we do not consider any variation in temperature throughout this paper, we omit the $T$ dependence in all following equations.

For the critical nucleus, the fact that we are at a saddle point in the free-energy landscape implies that the free-energy landscape is locally flat with respect to the two internal degrees of freedom.  Specifically,
\begin{eqnarray}
    \left(\frac{\partial \Omega_\text{tot}}{\partial V_X}\right)_{\mu, V, v} &=& 0 \label{eq:constraint1}, \\
    \left(\frac{\partial \Omega_\text{tot}}{\partial v}\right)_{\mu, V, V_X} &=& 0
    \label{eq:constraint2}, 
\end{eqnarray}
where the subscripts denote variables kept fixed during the derivation.
Notice that both these derivatives implicitly result in a change in the number of lattice sites $M$, as $v=V_X/M$.

In practice, it is helpful to define the \textit{interfacial free energy} $\gamma$ via
\begin{equation}
    \Omega_S(\mu,R,v) = \gamma(\mu,R, v) A, \label{eq:definegamma}
\end{equation}
with $A = 4\pi R^2$ the surface area of the crystal nucleus.
The constraints from Eqs. \ref{eq:constraint1} and \ref{eq:constraint2} result in
\begin{align}
    \label{eq:min_Vx}
    P_F + \omega_X + \frac{2\gamma}{R} + \left(\frac{\partial \gamma}{\partial R}\right)_{\mu,v} &= 0,\\
    \label{eq:min_v}
     \omega_X +P_X - \frac{3v}{R} \left(\frac{\partial \gamma}{\partial v}\right)_{\mu,R} &= 0.
\end{align}
Here $P_F$ and $P_X$ are, respectively, the pressures of the fluid and crystal phases far from the interface, and $\omega_X=\Omega_X/V_X$ represents the cost of increasing the size of the crystal nucleus while keeping the lattice spacing fixed. We will discuss the distinction between $P_X$ and $\omega_X$ in detail in the next subsection. 
Subtracting these two equations yields
\begin{eqnarray}\label{eq:pressure_diff1}
    P_X - P_F &=& \frac{2\gamma}{R} + \left(\frac{\partial \gamma}{\partial R}\right)_{\mu,v} + \frac{3v}{R} \left(\frac{\partial \gamma}{\partial v}\right)_{\mu,R} , \\
    &=& \frac{2 f}{R} + \left(\frac{\partial \gamma}{\partial R}\right)_{\mu,v},
\end{eqnarray}
where we have used the spherically averaged \textit{surface stress} $f$ defined as 
\begin{equation}
    f \equiv   \gamma + \frac{3v}{2} \left(\frac{\partial \gamma}{\partial v}\right)_{\mu,R}. \label{eq:f}
\end{equation}
Note that for a specific crystal facet, the surface stress is a tensor $f_{ij}$ given by the Shuttleworth equation \cite{shuttleworth1950surface, di2020shuttleworth}:
\begin{equation} \label{eq:shuttleworth}
    f_{ij} = \gamma \delta_{ij} + \partdev{\gamma}{\epsilon_{ij}},
\end{equation}
where  $\epsilon_{ij}$ the strain tensor associated with tangential deformations of the interface. Since we have assumed here that the interfacial free energy of our spherical nucleus can be described by an averaged $\gamma$, the surface stress similarly reduces to a single scalar quantity.

Importantly, the pressure difference cannot depend on the choice of dividing surface, so we are free to choose any dividing surface. A common and convenient choice is the so-called \textit{surface of tension} which satisfies:
\begin{equation}
   \left. \left( \frac{\partial \gamma}{\partial R}\right)_{\mu, v}\right|_{R=R^{*}} = 0,
\end{equation}
where the asterisk indicates variables evaluated under the condition that $R$ is chosen as the surface of tension \cite{}.
This results in the pressure difference
\begin{equation}\label{eq:pressure_diff}
    P_X - P_F = \frac{2\gamma^*}{R^*} + \frac{3v}{R^*} \left(\frac{\partial \gamma^*}{\partial v}\right)_{\mu,R}
    = \frac{2f^*}{R^*}.
\end{equation}
This equation tells us that the pressure inside the crystal nucleus is determined by the surface stress associated with the interface between the two phases. The two terms in the surface stress $f$ arise from the two effects that compressing the crystal nucleus has on the total surface free energy $\Omega_S$. First, similar to the gas--liquid case, compressing the nucleus results in a smaller surface area, reducing the interfacial free energy. Second, compressing a crystal nucleus changes its lattice spacing, which in turn may affect the interfacial free energy $\gamma$. Note that the latter effect can in principle be positive or negative.  For the simpler gas--liquid case, the latter term is always zero, so that in that case we recover the standard Young--Laplace equation: $P_\mathrm{liquid}-P_\mathrm{gas}=2\gamma^*/R^*$.

%%%%%%%%%%%%%%%%%%%%%%%%%%%%%%%%%%%%%%%%%%%%%%%%%%%%%%%%%%%%%%%

\subsubsection{Grand potential density $\omega_X$ versus the pressure $P_X$ for a crystal}

An important factor in the distinction between a fluid and a crystal is the relationship between the pressure and the grand potential density.
For a fluid, in the grand-canonical ensemble, we can write the grand potential simply as $\Omega_F(\mu, V_F)$, and the pressure can be obtained by taking the partial derivative with respect to $V_F$, while keeping $\mu$ constant:
\begin{equation}
    P_F = -\left(\frac{\partial \Omega_F}{\partial V_F}\right)_{\mu}. \label{eq:fluidP}
\end{equation}

Moreover, for any sufficiently large homogeneous system, we know that the grand potential is extensive meaning that we can write
\begin{eqnarray}
    \Omega_F(\mu, V_F) = V_F \omega_F(\mu), \label{eq:fluidomega}
\end{eqnarray}
where we have introduced the grand potential density $\omega_F$ of the fluid. Combining Eqs. \ref{eq:fluidP} and \ref{eq:fluidomega}, we trivially find:
\begin{equation}
\Omega_F = -P_F V_F.
\end{equation}

For a crystal, which can be under strain, the situation becomes more complicated. In the case where the only allowed strain is isotropic (as we assume in this paper), the grand potential of the crystal can be written as $\Omega_X(\mu, V_X, v)$, and is hence dependent on the unit cell volume $v = V_X/M$, with $M$ the number of lattice sites. Along the same lines as for the fluid, we again can write the pressure as
\begin{equation}
    P_X = -\left(\frac{\partial \Omega_X}{\partial V_X}\right)_{\mu,M}. \label{eq:crystalP}
\end{equation}
Note that here we additionally keep the number of lattice sites $M$ fixed. This is the pressure one would measure in standard equation-of-state calculations via computer simulations of bulk crystals. 

Interestingly, for the crystal, there is another derivative we could take which looks very strongly related. In particular, we could take the derivative of $\Omega_X$ with respect to the volume while keeping the lattice spacing fixed: $(\partial \Omega_X/\partial V_X)_{\mu,v}$.
%$\left(\frac{\partial \Omega_X}{\partial V_X}\right)_{\mu,v}$.
Note that while this change may look very small, the physics of this variation is quite different from the one in Eq. \ref{eq:crystalP}. In particular, while previously we were deforming the crystal during our compression or extension of the system, we now are simply changing the amount of crystal in our system, scaling the number of lattice sites, volume, and particles all proportionally to each other. For this transformation, we once again recover extensivity for a large enough system. In other words:
\begin{eqnarray}
    \Omega_X(\mu, V_X,v) = V_X \omega_X(\mu, v).\label{eq:crystalomega}
\end{eqnarray}
From this we obtain:
\begin{equation} \label{eq:crystaldOmegadV}
    \left(\frac{\partial \Omega_X}{\partial V_X}\right)_{\mu,v} = \omega_X(\mu,v).
\end{equation}
However, given the differences between the derivatives in Eqs. \ref{eq:crystalP} and \ref{eq:crystaldOmegadV}, this means that $\Omega_X \neq -P_X V_X$ in general.

To see the relationship between $P_X$ and $\omega_X$, we simply use the chain rule:
\begin{eqnarray}
    P_X &=& -\left(\frac{\partial \Omega_X}{\partial V_X}\right)_{\mu,M} \\
    &=& -\left(\frac{\partial \Omega_X}{\partial V_X}\right)_{\mu,v} -
    \left(\frac{\partial \Omega_X}{\partial v}\right)_{\mu,V_X}
    \left(\frac{\partial v}{\partial V_X}\right)_{\mu,M}\\
    &=& -\omega_X(\mu,v) -
    \left(\frac{\partial \Omega_X}{\partial v}\right)_{\mu,V_X}
    \frac{1}{M}.  \label{eq:Pcrystalproper}
\end{eqnarray}
In the special case of an equilibrium crystal, where the lattice spacing $v$ has been optimized to correspond to the minimum in the free energy $\Omega_X$, the last term in this expression vanishes. Hence, $P_X^\mathrm{eq} = -\omega_X$, and we once again recover $\Omega_X = -P_X^{\mathrm{eq}} V_X$.

However, if the crystal is under any strain, then by definition the derivative in the second term in Eq.  \ref{eq:Pcrystalproper} will be non-zero, and hence $\omega_X \neq - P_X$ (and $\Omega_X \neq - P_X V_X$) in general. 
Since our nucleus is under strain due to the presence of an interface, $\omega_X$ cannot be trivially identified with the mechanical pressure in the interior of the nucleus. Instead, it is a grand potential density, which we can regard as a perturbation of the grand potential of an unstrained crystal at the same chemical potential.
To this end, we make the approximation that the crystal phase inside the nucleus is not strongly distorted with respect to its equilibrium lattice spacing $v^\mathrm{eq}(\mu)$, and expand around equilibrium leading to:
\begin{eqnarray}
    \omega_X(\mu, v) &=& \omega_X(\mu, v^\mathrm{eq}) + \left.\diffix{\omega_X}{v}{\mu}\right|_{v=v^\mathrm{eq}} (v-v^\mathrm{eq}) \nonumber\\ && + \, \mathcal{O}((v-v^\mathrm{eq})^2) \\
    &=& -P_X^\mathrm{eq}(\mu) + \mathcal{O}((v-v^\mathrm{eq})^2). \label{eq:Pthermodynamic}
\end{eqnarray}
Here, we have used the fact that in equilibrium, the grand potential is minimized with respect to $v$, and equal to $\Omega_X = -P_X^\mathrm{eq}(\mu)V_X$.  In other words, to linear order in the strain on the crystal, $\omega_X$ is given by (minus) the pressure of the unstrained equilibrium crystal at the chemical potential of the fluid. This quantity $P_X^\mathrm{eq}$ corresponds exactly to what is referred to as the thermodynamical pressure in Ref. \onlinecite{montero2020young}.

Combining Eq. \ref{eq:Pthermodynamic} with Eqs. \ref{eq:min_Vx} and \ref{eq:min_v}, we get the relation
\begin{align}
     P_X^\mathrm{eq}(\mu) - P_F &= \frac{2\gamma}{R} + \left(\frac{\partial \gamma}{\partial R}\right)_{\mu,v}.\label{eq:YoungLaplacePeq}
\end{align}
Note that at the surface of tension, the last term in Eq.  \ref{eq:YoungLaplacePeq} can again be eliminated:
\begin{align}
     P_X^\mathrm{eq}(\mu) - P_F &= \frac{2\gamma^*}{R^*}.\label{eq:YoungLaplacePeq2}
\end{align}
This recovers the Young--Laplace-like equation presented in Ref. \onlinecite{montero2020young}, where the authors used a framework which uses as reference for the crystal phase inside the nucleus a bulk equilibrium crystal phase with the same chemical potential as the fluid. In contrast, in this work we consider the actual crystal phase inside the nucleus.

In the remainder of this section, we examine the pressure difference directly in simulations of equilibrium crystal nuclei of hard spheres.

%%%%%%%%%%%%%%%%%%%%%%%%%%%%%%%%%%%%%%%%%%%%%%%%%%%%%%%%%%%%%%%

\begin{figure*}[t!]
    \includegraphics[width=0.95\linewidth]{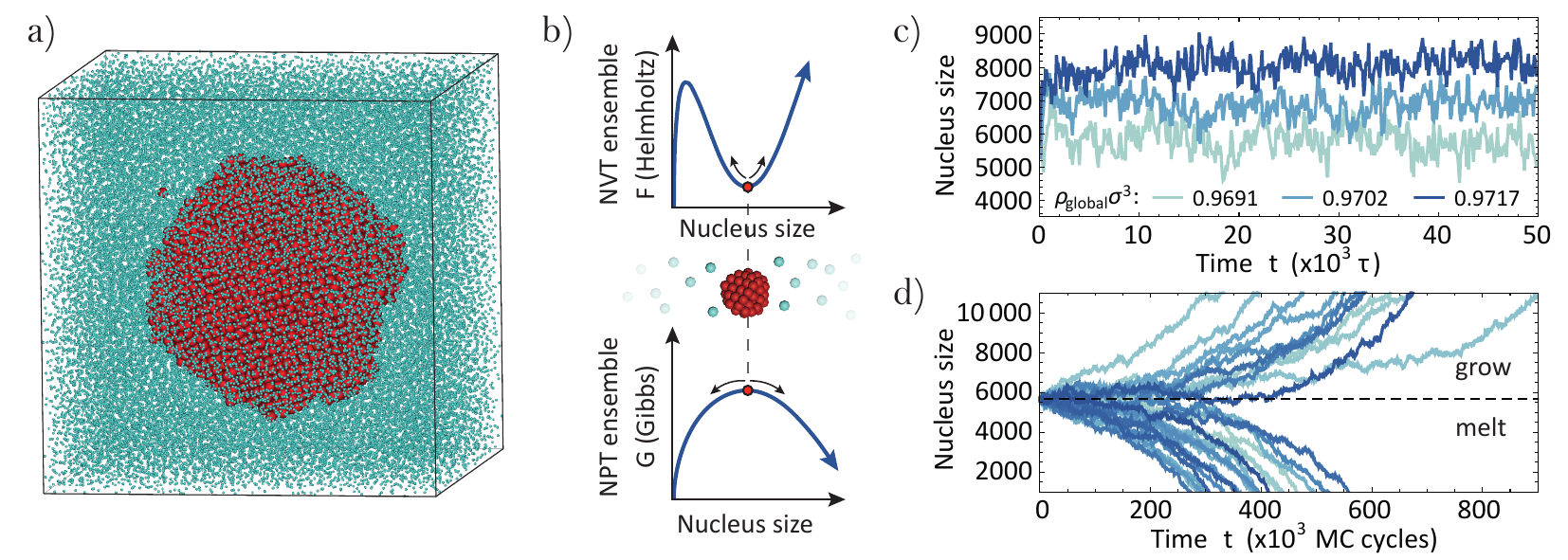} 
    \caption{\label{fig:cluster} 
    \textbf{a)} Snapshot of nucleus A after the overlaps are removed. Particles classified as crystal are depicted in red, whereas particles classified as fluid are depicted in blue and at a quarter of their actual size to make the crystal nucleus visible. \textbf{b)} Sketch of the Helmholtz free energy of the $NVT$ ensemble and the Gibbs free energy of the $NPT$ ensemble as a function of nucleus size. Note that the stable nucleus in the $NVT$ ensemble corresponds to a critical nucleus in the $NPT$ ensemble. For the $\mu V T$ ensemble, the free-energy difference is the same as in the $NPT$ ensemble (not shown). \textbf{c)} Nucleus size during an EDMD simulation in the $NVT$ ensemble of nucleus A at three different global densities. \textbf{d)} Nucleus size during 30 MC simulations in the $NPT$ ensemble, all started from an equilibrated configuration of nucleus A ($\rho_\text{global}\sigma^3=0.9691$, $\beta P_\text{global}\sigma^3=12.5945$). %eta=0.5074
    Note that, in terms of long-time diffusion time of the fluid particles, the $NVT$ EDMD simulations of \textbf{c)} ran approximately 50 times longer than the $NPT$ MC simulations of \textbf{d)}.
    }
\end{figure*}

\subsection{Methods} \label{ssec:pressuremethods}

In order to examine stable configurations containing a spherical nucleus, we simulate our systems in the $NVT$ ensemble.
In this ensemble, the free-energy landscape can exhibit a local minimum corresponding to a spherical nucleus, as sketched in Fig. \ref{fig:cluster}b. This free-energy minimum corresponds (via a Legendre transform) to a saddle point (i.e. a critical nucleus) in the free-energy landscape in both the $NPT$ and $\mu VT$ ensembles \cite{richard2018crystallizationII,yang1985thermodynamical,rusanov1964thermodynamics} and is the saddle point discussed in the Theory section above. 
We use event-driven molecular dynamics (EDMD) simulations \cite{alder1959studies,smallenburg2022efficient} to simulate systems of perfectly hard spheres with diameter $\sigma$ and mass $m$. We do not make use of a thermostat, and hence the total energy of the system (which consists only of the kinetic energy) is fixed. This in turn also fixes the temperature $T$. The time unit of our simulations is given by $\tau=\sqrt{\beta m \sigma^2}$, where $\beta=1/k_BT$, with $k_B$ the Boltzmann constant.

As initial configurations for our system, we use two different approaches.  First, we use configurations from the study in Refs. \onlinecite{montero2020interfacial} and \onlinecite{montero2020young}. These are equilibrated nuclei of pseudo-hard spheres (PHS), i.e. spheres interacting via a nearly-hard pair potential \cite{jover2012pseudo}. Specifically, we make use of the configurations labeled IV and V in Refs. \onlinecite{montero2020interfacial} and \onlinecite{montero2020young}, which we here label A and B, respectively. 
We turn these configurations into pure hard-sphere configurations by replacing the particles with hard spheres with a diameter slightly smaller than $\sigma$, such that there are no overlaps, and then rapidly growing these particles back to the full diameter $\sigma$ using the implementation of the Lubachevsky-Stillinger approach \cite{lubachevsky1990geometric} of Ref. \onlinecite{smallenburg2022efficient}.
Note that these growth simulations are extremely short (taking less than $0.05\tau$), such that the particles do not rearrange significantly and the overall size of the nucleus does not change significantly during this step. 
As a second source of initial configurations we generate new coexisting states by initializing systems of different numbers of particles $N$ and volumes $V$ in a fully crystalline state, surrounded by a thin layer of empty space on all sides. After equilibration, this results in a spherical nucleus whose size is determined by the initial $N$ and $V$. Overall, this results in a set of initial configurations (labeled C throughout this paper) with sizes spanning from $N=2\cdot10^4$ to $3\cdot10^5$ particles, and with nuclei typically covering on the order of $15\%$ of the box volume.

In the EDMD simulations, the global pressure can be easily calculated from the momentum transfer during collisions \cite{alder1960studies}, i.e.
\begin{equation}\label{eq:pressure}
    \beta P_{kl}/\rho = 1 +  
    \frac{\sum_\text{collisions} \Pi_{ij}^{kl}}{ N \Delta t},
\end{equation}
where $P_{kl}$ indicates the $kl$-component of the pressure tensor, $\rho=N/V$ is the number density, $\sum_\text{collisions}$ indicates the sum over all collisions during a time interval $\Delta t$, and
$\Pi_{ij}^{kl}$ indicates $kl$-component of the momentum transfer during a collision between particles $i$ and $j$. 
For monodisperse hard spheres, $\Pi_{ij}^{kl}=-m {v}_{ij}^{k} {r}_{ij}^{l}$. Here 
${r}_{ij}^{l}$ indicates the $l$-th component of the center-to-center distance vector $\mathbf{r}_{ij}=\mathbf{r}_{j}-\mathbf{r}_{i}$ at collision, and ${v}_{ij}^{k}$ indicates the $k$-th component of  $\mathbf{v}_{ij}=\mathbf{v}_{j}-\mathbf{v}_{i}$, with $\mathbf{v}_{i(j)}$ the velocity of particle $i$ ($j$) before collision. 
Note that in (hydrostatic) equilibrium the pressure tensor will average to $P_{ij} = P\, \delta_{ij}$, with $P$ the total pressure and $\delta_{ij}$ the Kronecker delta. Consequently, $P$ is obtained by taking one-third of the trace of the pressure tensor.

In addition to the global pressure, we are also interested in measuring the pressure profile as a function of the radial distance to the center of the nucleus. 
To this end, we divide the system into spherical shells around the center-of-mass of the nucleus, and keep track of the momentum transfer inside each shell, as well as the local density \cite{}. 
For the radial profile of the total pressure, one can make the reasonable approximation that, for each collision, half of the momentum transfer is added to the shell in which $\mathbf{r}_i$ lies and half to the shell in which $\mathbf{r}_j$ lies \cite{}.
This approximation, however, is not valid for the components of the momentum transfer normal and tangential to the crystal--fluid interface (we have need for them in Section \ref{sec:stress}).
Hence, to obtain the total, normal, and tangential pressure profiles, we instead use the method described in Refs. \onlinecite{nakamura2011novel} and \onlinecite{nakamura2015precise}.
In order to explain this method, consider a collision between particles $i$ and $j$ and define the straight path from $\mathbf{r}_i$ to $\mathbf{r}_j$ as $\boldsymbol{\ell}(\lambda) = \mathbf{r}_i + \lambda \mathbf{r}_{ij}$ with $0\leq\lambda\leq1$.
The method then assigns a fraction of the momentum transferred during this collision to each of the shells traversed by $\boldsymbol{\ell}$ based on the part of $\boldsymbol{\ell}$ inside the shell. We indicate the part of $\boldsymbol{\ell}$ inside a certain shell by $\boldsymbol{\ell}_a\equiv\boldsymbol{\ell}(\lambda_a)$ and $\boldsymbol{\ell}_b\equiv\boldsymbol{\ell}(\lambda_b)$, which both mark either an intersection with the shell boundary or a terminal point of $\boldsymbol{\ell}$ in the shell (i.e. $\lambda_a=0$ or $\lambda_b=1$).
The total, normal, and tangential contributions of the momentum transfer of the collision to that shell are then given by
\begin{align}
    \left[\Pi_{ij} \right]^b_a &= -(\mathbf{v}_{ij}\cdot\mathbf{r}_{ij}) \; \frac{\alpha_b-\alpha_a}{3|\mathbf{r}_{ij}|^2}, \\
    \left[\Pi_{ij}^\perp \right]^b_a &= -(\mathbf{v}_{ij}\cdot\mathbf{r}_{ij}) \; \frac{\alpha_b-\alpha_a}{|\mathbf{r}_{ij}|^2} - \left[ G_{ij} \right]^b_a , \\
    \left[\Pi_{ij}^\parallel \right]^b_a &= \frac{1}{2} \left[ G_{ij} \right]^b_a , 
\end{align}
where
\begin{align}
    \left[ G_{ij} \right]^b_a  &= \frac{-(\mathbf{v}_{ij}\cdot\mathbf{r}_{ij}) |\boldsymbol{\omega}| }{|\mathbf{r}_{ij}|^2} \left[\arctan\frac{\alpha_b}{|\boldsymbol{\omega}|} -\arctan\frac{\alpha_a}{|\boldsymbol{\omega}|} \right]  \nonumber \\
    &+ \frac{-(\mathbf{v}_{ij}\times\mathbf{r}_{ij}) \cdot \boldsymbol{\omega} }{|\mathbf{r}_{ij}|^2} \; \ln\frac{|\boldsymbol{\ell}_b|}{|\boldsymbol{\ell}_a|} . 
\end{align}
Here, we introduced the variables $\boldsymbol{\omega} = \mathbf{r}_{ij}\times\mathbf{r}_i$ and $\alpha(\lambda) = \mathbf{r}_{i}\cdot\boldsymbol{\ell}(\lambda)$, and we defined $\alpha_a\equiv\alpha(\lambda_a)$ and $\alpha_b\equiv\alpha(\lambda_b)$. The list of values for $\alpha_{a(b)}$ and $|\boldsymbol{\ell}_{a(b)}|$ for each collision is easy to compute (see Ref. \onlinecite{nakamura2011novel}).
Note that the normal and tangential pressure profiles can also be obtained in the post analysis from the radial profile of the total pressure, see Supplementary Material.

In order to keep track of the size and center of mass of the nucleus during our simulations, we classify each particle in the system as either fluid or crystal using the 6-fold Ten Wolde bonds \cite{tenwolde1996simulation}
\begin{equation}\label{eq:d6}
    d_6(i,j) = \frac{ \sum_m q_{6m}(i) q_{6m}^\dagger(j) }{ \sqrt{ \left(\sum_m |q_{6m}(i)|^2 \right) \left(\sum_m |q_{6m}(j)|^2 \right) } },
\end{equation}
where $^\dagger$ indicates the complex conjugate, $\sum_m$ indicates the sum over $m\in[-6,6]$, and $q_{6m}$ are Steinhardt's 6-fold bond-orientational order parameters \cite{steinhardt1983bond}. Particle $i$ is classified as crystal if it has 9 or more neighboring particles $j$ with which it has a crystal-like bond, i.e. $d_6(i,j)>0.7$. The neighbors of particle $i$ are defined as all particles $j$ with $|\mathbf{r}_{ij}|<1.45\sigma$, which, for all systems studied, roughly corresponds to the first minimum of the radial distribution function.

%%%%%%%%%%%%%%%%%%%%%%%%%%%%%%%%%%%%%%%%%%%%%%%%%%%%%%%%%%%%%%%

\subsection{Results} \label{ssec:pressureresult}

We begin our investigation by equilibrating the initial configurations of nuclei. A sample initial configuration is shown in Fig. \ref{fig:cluster}a.
The nuclei were equilibrated for $5\cdot10^4\tau$. 
As the coexistence region of pseudo-hard spheres is slightly different than that of hard spheres\cite{espinosa2013fluid}, simulating the nuclei from sets A and B at their original global density resulted in a noticeable increase of the nucleus size, with nucleus B even becoming system spanning. To address this, we equilibrate these nuclei for a small range of global densities, slightly lower than that of the original pseudo-hard-sphere configurations. Figure \ref{fig:cluster}c shows the size of the crystal nucleus during this equilibration run for nucleus A at three different global densities. One can see that the nuclei equilibrate quickly.

To confirm that the equilibrated nuclei, which are stable in the $NVT$ ensemble, are critical nuclei in the $NPT$ ensemble (see Fig. \ref{fig:cluster}b), we take a few configurations and start 30 Monte Carlo (MC) simulations in the $NPT$ ensemble from each configuration \cite{frenkelbook}. For the pressure in these MC simulations, we use the average global pressure measured during the EDMD simulation. To make the sampling in the MC simulations more efficient we implemented Almarza's algorithm for the volume changes \cite{almarza2007cluster}. We indeed observe that each nucleus melts or grows with a roughly 50/50 probability, see Fig. \ref{fig:cluster}d.

\begin{figure}[t!]
\begin{tabular}{l}
     a) \\[-0.4cm]
     \includegraphics[width=\figwidthA]{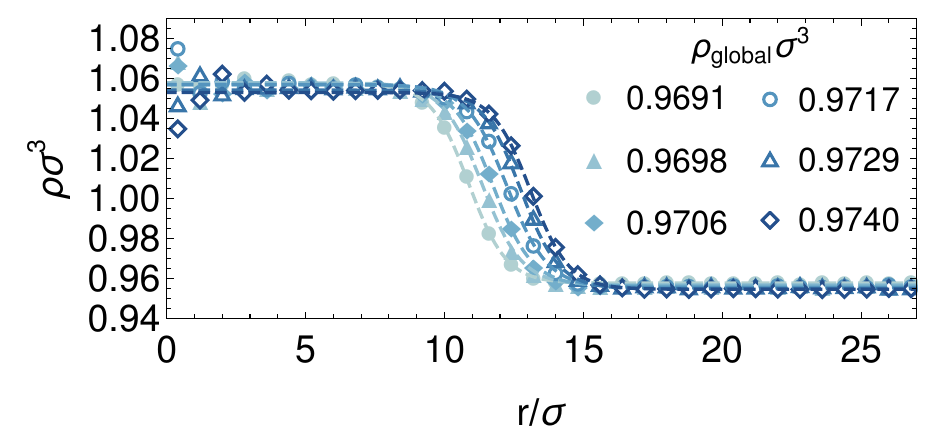} \\
     b) \\[-0.4cm]
    \includegraphics[width=\figwidthA]{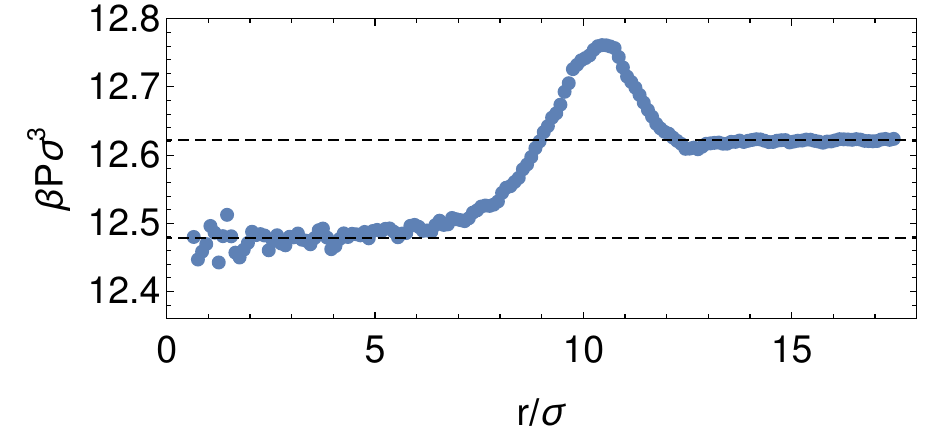} 
\end{tabular}
    \caption{\label{fig:profile} 
    \textbf{a)} The radial density profile for nucleus A (closed markers) and nucleus B (open markers) for a few different global densities. For clarity, all profiles are displayed with a bin width of $0.8\sigma$ and the dashed lines are guides to the eye. 
    \textbf{b)} The radial pressure profile for nucleus A at $\rho_\text{global}\sigma^3=0.9691$. The dashed lines indicate the average pressures in the ``bulk'' crystal and fluid phases. 
    }
\end{figure}

\begin{figure}[t]
    %\centering
    \includegraphics[width=\figwidthA]{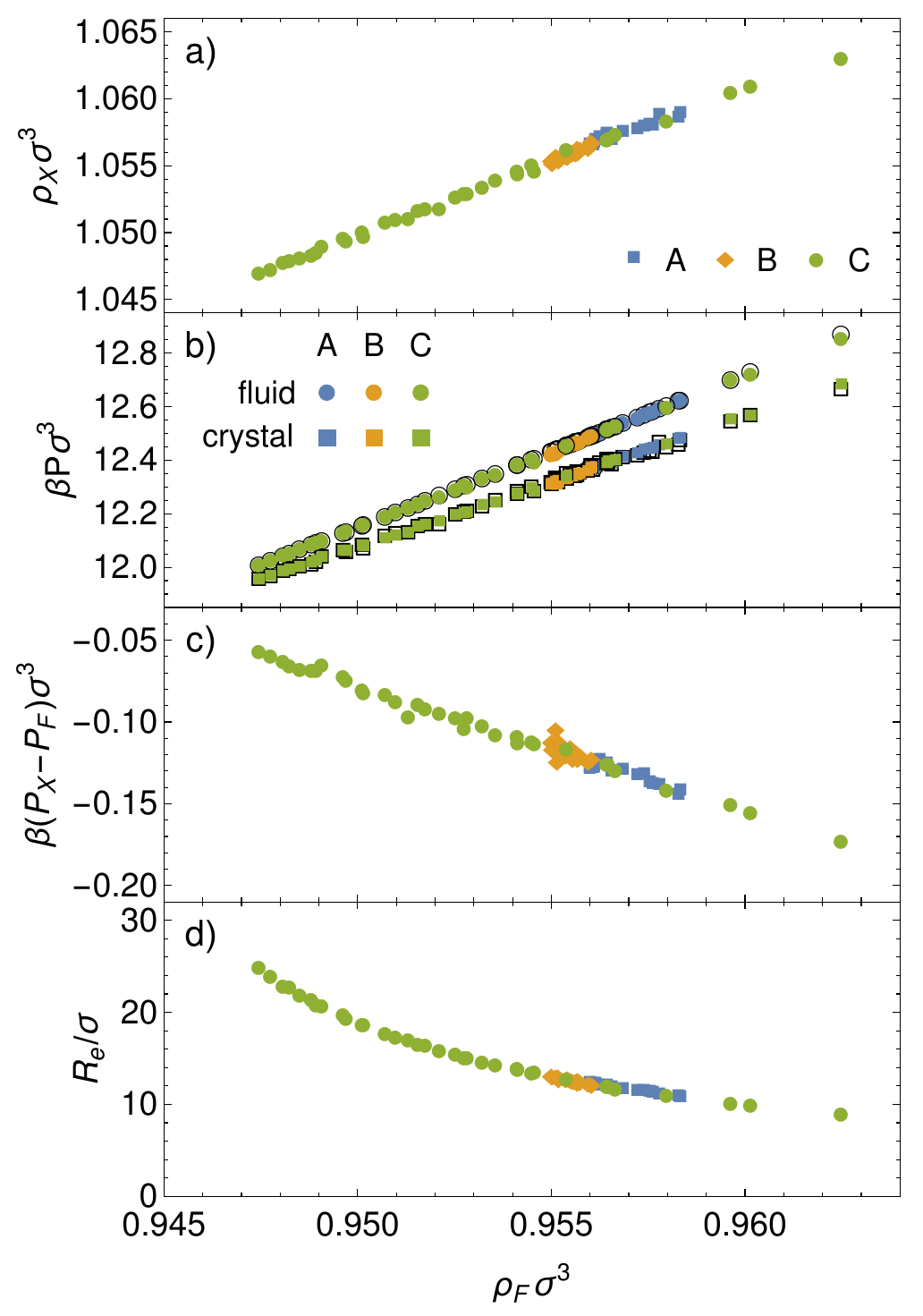}
    \caption{\label{fig:bulkvalues} 
    Different thermodynamic properties of the crystal nucleus and surrounding fluid for all investigated nuclei, all as a function of the density of the fluid phase.
    \textbf{a)} The density inside the crystal nucleus. \textbf{b)} The measured pressures (closed, colored markers), as well as the pressures obtained from the equation of states evaluated at the measured densities (black, open markers). \textbf{c)} The pressure difference between the crystal nucleus and the surrounding fluid. \textbf{d)} The equimolar radius $R_e$ of the nucleus.
    }
\end{figure}

Next, switching back to the canonical ensemble, we measure the radial density and pressure profiles around the center of the nucleus. For this we take the equilibrated configurations of the EDMD simulations, and start new EDMD simulations of $10^4\tau$ in total. Since the center of mass of the nucleus slowly drifts during the simulation, we update it each $0.5\tau$. During this update we also measure the number of particles in each spherical shell around the center of mass. We use a bin width of $0.1\sigma$ for the spherical shells. In Fig. \ref{fig:profile}, we show a selection of the resulting density profiles, as well as a typical pressure profile. From the density profiles, one can clearly see that the nucleus grows with increasing global density, as was predicted for PHS in Ref. \onlinecite{montero2022thermodynamics}.
Furthermore, looking at the pressure profile, we see that the pressure both inside and outside of the nucleus reach a well-defined value far away from the interface. 
For the fluid, this pressure corresponds to the average global pressure of the entire system, which is due to the mechanical equilibrium condition \cite{montero2020young}.
Importantly, we find a lower pressure inside the crystal nucleus than in the surrounding fluid, consistent with what was observed in Ref. \onlinecite{montero2020young} for pseudo-hard spheres.

From the plateau values of the density and pressure profiles, we directly obtain the densities and pressures of both phases. In Fig. \ref{fig:bulkvalues}a we plot the crystal density as a function of the density of the fluid phase. Note that the density of both the fluid and the crystal (Fig. \ref{fig:bulkvalues}a) are always significantly above the freezing and melting densities, i.e. $\rho_F^\text{coex}\sigma^3=0.93918(1)$ and $\rho_X^\text{coex}\sigma^3=1.0375(3)$ \cite{frenkelbook,polson2000finite,smallenburg2024simple}. This is consistent with the idea that these nuclei are critical, which can only occur in supersaturated fluids.

In Figs. \ref{fig:bulkvalues}b-c, we explore the pressure difference between the two phases, and indeed find that for the wide range of nucleus sizes studied, the pressure inside the nucleus is always lower than that of the fluid. The absolute pressure difference $|P_X - P_F|$ gradually decreases with decreasing $\rho_F$, consistent with the requirement that it vanishes at the freezing density, where the two phases should have equal pressures. To double-check our pressure measurements, we additionally plot in Fig. \ref{fig:bulkvalues}b the pressures as obtained from the hard-sphere equations of state of both phases \cite{kolafa2004accurate,speedy1998pressure}, evaluated at the measured densities (black open symbols), and find excellent agreement, similar to what was seen for PHS\cite{montero2020young}. 
Finally, in Fig. \ref{fig:bulkvalues}d we show the equimolar radius $R_e$ of the nucleus, calculated using Eq. \ref{eq:N} (with $N_S = 0$). Clearly we obtain nuclei spanning a wide range of sizes. Note, however, that all of these simulations are still at relatively low supersaturation. In fact, although the range of fluid pressures obtained here ($12.0<\beta P_F\sigma^3<12.9$) is certainly above the coexistence pressure $\beta P_\text{coex}\sigma^3=11.5646(5)$\cite{smallenburg2024simple}, it is much below the pressures where spontaneous nucleation can be feasibly studied using brute-force simulations (typically above pressures $\beta P_F \sigma^3 \gtrsim 15$ \cite{filion2010crystal, wohler2022hard, gispen2023brute}). 
The smaller nuclei at higher supersaturation are harder to stabilize in unbiased $NVT$ simulations. The free-energy well associated with these nuclei becomes less deep for smaller system sizes, making them more susceptible to escape via thermal fluctuations. As a result, small nuclei either melt or grow out, spanning the box.

We note that all nuclei investigated are nuclei of the face-centered cubic (FCC) crystal. In the Supplementary Material, we show that the thermodynamic properties of nuclei of the hexagonal close-packed (HCP) crystal agree with those of FCC nuclei.

From Eq. \ref{eq:pressure_diff}, we see that the lower pressure inside the crystal phase should be linked to a negative surface stress $f^*$ of the interface between the fluid and the crystal. In particular, the negative pressure differences and nucleus radii shown in Fig. \ref{fig:bulkvalues}c and \ref{fig:bulkvalues}d correspond to a surface stress of approximately $f^* \simeq -0.7k_BT/\sigma^2$ (assuming for the moment that the equimolar radius $R_e \simeq R^*$). 
In the next section, we examine the surface stress of hard spheres in more detail.

%%%%%%%%%%%%%%%%%%%%%%%%%%%%%%%%%%%%%%%%%%%%%%%%%%%%%%%%%%%%%%%
%%%%%%%%%%%%%%%%%%%%%%%%%%%%%%%%%%%%%%%%%%%%%%%%%%%%%%%%%%%%%%%

\section{Surface stress of a spherical nucleus}\label{sec:stress}

In this section, we examine the surface stresses associated with a fluid--crystal interface in hard spheres in more detail. We begin by revisiting the theory associated with the surface stress for flat crystal--fluid interfaces, and then extend this to spherical nuclei. Using our measurements of the pressure profiles described in the previous section, we then determine the surface stress for a spherical nucleus of hard spheres as a function of the metastable fluid density.

%%%%%%%%%%%%%%%%%%%%%%%%%%%%%%%%%%%%%%%%%%%%%%%%%%%%%%%%%%%%%%%

\subsection{Theory}

As a starting point, we consider a fluid--crystal coexistence with a flat interface (i.e. in a slab geometry), in the grand-canonical ensemble. Specifically, we consider a periodic simulation box elongated along the $z$-axis containing two interfaces perpendicular to the long axis of the box. Note that for a monodisperse system in the grand-canonical ensemble, such a configuration is metastable and corresponds to a saddle point in the free energy.  

In this geometry, the lattice spacing of the crystal is imposed by the periodicity of the system along the $x$ and $y$ axes of the box. Specifically, the lattice spacing $a_{x(y)} = L_{x(y)}/M_{x(y)}$, where $M_{x(y)}$ is the number of lattice sites along the ${x(y)}$-direction, which we will keep fixed in this entire derivation. Along the longer $z$-axis, both the crystal lattice spacing and the number of crystalline layers can fluctuate.
For such a system, the grand potential can be written as
\begin{eqnarray}
    \Omega_\mathrm{tot}(\mu,V,L_x,L_y,L_z;V_X,a_z) &=& 
    \Omega_F(\mu,V_F)  \nonumber\\ 
    &&+
    \Omega_X(\mu,V_X,a_z,L_x,L_y) \nonumber\\
    &&+  
    2 \gamma(\mu,a_x,a_y) A. \label{eq:Fcoex}
\end{eqnarray}
Here, $a_i$ is the lattice spacing of the crystal in the $i$-direction, $A$ is the surface area of one of the interfaces, and the factor 2 arises from the presence of two interfaces. Note that for flat interfaces, the interfacial free energy and surface stress are independent of the choice of dividing surface. 

Minimization of $\Omega_\mathrm{tot}$ with respect to $V_X$ and $a_z$ gives
\begin{eqnarray}
    \diffix{\Omega_\mathrm{tot}}{V_X}{a_z} = P^{zz}_F + \omega_X = 0 ,\\
    \diffix{\Omega_\mathrm{tot}}{a_z}{V_X} = -P^{zz}_X - \omega_X = 0 ,
\end{eqnarray}
which leads to
\begin{eqnarray}
    P^{zz}_F &=& P^{zz}_X.
\end{eqnarray}
Here, $P^{zz}$ denotes the $zz$-component of the pressure tensor $P^{ij}$ of a given phase, which can be anisotropic for a crystal under strain or a system containing an interface.  Additionally, we note the pressure tensor inside the fluid phase is necessarily isotropic ($P^{zz}_F = P_F$), and that under equilibrium coexistence conditions, the pressure tensor \textit{inside} the crystal phase must also be isotropic: $P^{xx}_X = P^{yy}_X = P^{zz}_X = P_X = P_F$ (see e.g. Ref. \onlinecite{smallenburg2024simple}). Hence, the only anisotropic contribution to the global pressure tensor of a system under equilibrium coexistence conditions comes from the interface.

We can now consider how the free energy changes upon applying an infinitesimal elongation of the system along the $x$-axis, i.e. tangential to the interface. From the definition of the pressure tensor, we can write
\begin{eqnarray}
    \diffix{\Omega_\mathrm{tot}}{L_x}{\mu,V,L_y,L_z} &=& -P_{xx} L_y L_z. \label{eq:Pxx1}
\end{eqnarray}
Using Eq. \ref{eq:Fcoex}, we can also decompose this free-energy change into contributions arising from the fluid, crystal, and interface. 
For elongation along the $x$-axis, we obtain
\begin{eqnarray}
   \diffix{\Omega_\mathrm{tot}}{L_x}{\mu,V,L_y,L_z} 
   &=& -\frac{V_F}{V} P_F L_y L_z
     -\frac{V_X}{V} P_X L_y L_z \nonumber\\
    && +2 \gamma L_y + 2\partdev{\gamma}{L_x} A  \nonumber\\
    &=& -P_F L_y L_z + 2L_y\left(\gamma + \partdev{\gamma}{\epsilon_{xx}}\right) \nonumber\\
    &=&-P_{zz}{L_y L_z} + 2L_y f_{xx}, \label{eq:Pxx2}
\end{eqnarray}
where $\epsilon_{xx}$ is the applied strain on the interface along the $x$-direction, and we used the Shuttleworth equation (Eq. \ref{eq:shuttleworth}) in the last step.
Combining Eqs. \ref{eq:Pxx1} and \ref{eq:Pxx2}, we obtain:
\begin{equation}
    f_{xx} = \frac{1}{2} L_z(P_{zz} - P_{xx}). \label{eq:fflat} 
\end{equation}
The analogous expression for $f_{yy}$ can be derived in the same way. This provides us with a method to directly determine the surface stress for a flat interface from direct coexistence simulations. Note that $f$ can be seen as a tensor of elastic constants for the crystal--fluid interface, and obeys the same symmetry considerations as the elastic tensor of a two-dimensional solid. Hence, when the crystal plane facing the fluid has square or hexagonal symmetry, we expect that $f_{xx} = f_{yy}$. 

It is important to note that, similar to the interfacial free energy, the surface stress is expected to depend on which crystal plane faces the fluid. In the case of a spherical nucleus, the relevant values of $\gamma$ and $f$ correspond to their spherically averaged values, taken over the entire nucleus surface. In principle, one could estimate the surface stress of a spherical nucleus by measuring it for a number of different crystal planes and taking an average (as has been done for $\gamma$, see e.g. Refs. \onlinecite{hartel2012tension, davidchack2010hard}). However, a more direct measure could be obtained by instead extending the approach for flat interfaces to spherical nuclei.
To this end, we consider a system containing a spherical nucleus in the grand-canonical ensemble, and measure the normal and tangential components of the pressure as a function of the radial distance from the center of the nucleus.

We would now like to deform the interface in a way that stretches its surface area uniformly, while keeping the spherical geometry and the radius of curvature of the surface fixed. This is not physically possible in a spherical nucleus, but we can imagine performing this deformation only locally, on a narrow cone-shaped subvolume of the system, with its tip located at the center of the nucleus, and extending radially outward into the fluid phase up to a maximum distance $R_\mathrm{max}$ (see e.g. Ref. \onlinecite{rowlinson2013molecular}). 
We now consider changing the volume of this cone by modifying only its opening angle $\theta$, and examine the effect on the grand potential. Looking at this deformation in spherical coordinates, this moves the boundaries of the cone-shaped region outward in the direction tangential to the interface. The change in free energy due to this deformation can be written in terms of the total pressure exerted on the sides of the conical volume:
\begin{equation}
\partdev{\Omega_\mathrm{tot}}{\theta} = -2\pi \sin \theta \int_0^{R_\mathrm{max}} \dd r r^2 P_\parallel(r),
\end{equation}
where $P_\parallel(r)$ is the tangential pressure profile as a function of the distance to the center of the nucleus. This can be measured directly in simulations similar to how we measured the total pressure profile \cite{nakamura2011novel, nakamura2015precise}, see Section \ref{ssec:pressuremethods} for the details.

Alternatively, by splitting the free energy up into contributions from the fluid, crystal, and interface, we can write the same derivative as
\begin{eqnarray}
    \partdev{\Omega_\mathrm{tot}}{\theta} &=&  \diffix{\Omega_F}{\theta}{}
    + \diffix{\Omega_X}{\theta}{} + \diffix{\gamma A}{\theta}{}\\
    &=& \frac{2\pi \sin\theta}{3} \left(-P_F (R_\mathrm{max}^3 - R^3) -P_X R^3 + 3f R^2\right). \nonumber\\
\end{eqnarray}
Combining the two expressions, we can measure $f$ by calculating:
\begin{eqnarray}
 f &=& \int_0^{R_\mathrm{max}} \dd r \frac{r^2}{R^2} \left[P_\mathrm{step}(r) - P_\parallel(r)\right], \label{eq:fspherical}
\end{eqnarray}
where $P_\mathrm{step}(r)$ is a step function based on our choice for the radius of the dividing surface $R$ ($P_\mathrm{step}(r<R) = P_X$, $P_\mathrm{step}(r>R) = P_F$).

%%%%%%%%%%%%%%%%%%%%%%%%%%%%%%%%%%%%%%%%%%%%%%%%%%%%%%%%%%%%%%%

\subsection{Results}

We first measure the surface stress for planar fluid--crystal interfaces using direct-coexistence simulations. To set up an equilibrium coexistence between a fluid and an unstrained crystal, we follow the approach of Ref. \onlinecite{smallenburg2024simple}. We perform EDMD simulations of hard spheres in an elongated simulation box, where the initial configuration is mostly filled with a perfect FCC crystal at a chosen initial density $\rho_\text{init}$. The global density of the simulation box $\rho < \rho_\text{init}$ is set by introducing a slab of empty space in the simulation box, oriented parallel to the long axis of the box. We fix this overall density to be inside the fluid--crystal coexistence region ($\rho \sigma^3 = 0.99$), such that during equilibration the crystal slab partially melts, resulting in a system where the fluid and crystal phases coexist with each filling roughly half of the simulation box. The crystal is oriented such that either the (100) plane or the (111) plane faces the fluid, with the configurations containing $N=14850$ and $N=16133$ particles, respectively. In this geometry, the lattice spacing normal to the interface can relax and adapt to the pressure, while the lattice spacing in the two tangential directions is fixed by the periodic boundary conditions, and hence by $\rho_\text{init}$. After equilibration, we measure the pressure tensor.

\begin{figure}[t!]
\begin{tabular}{l}
     a) \\[-0.4cm]
     \includegraphics[width=\figwidthA]{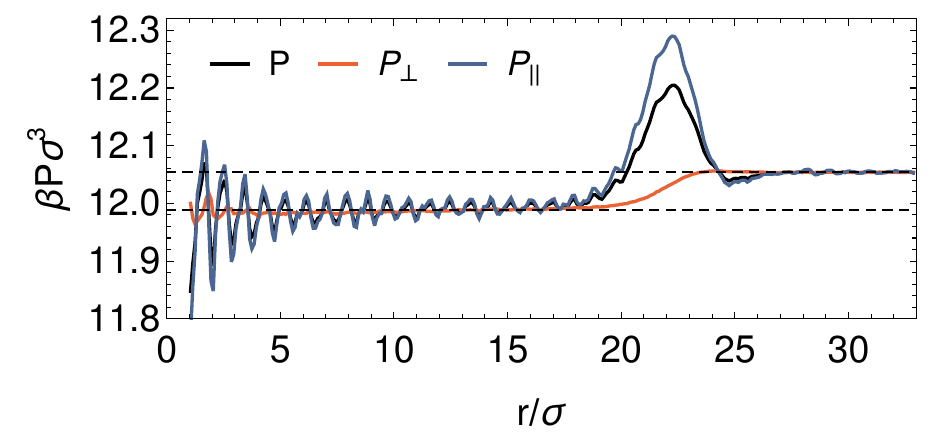} \\
     b) \\[-0.4cm]
    \includegraphics[width=\figwidthA]{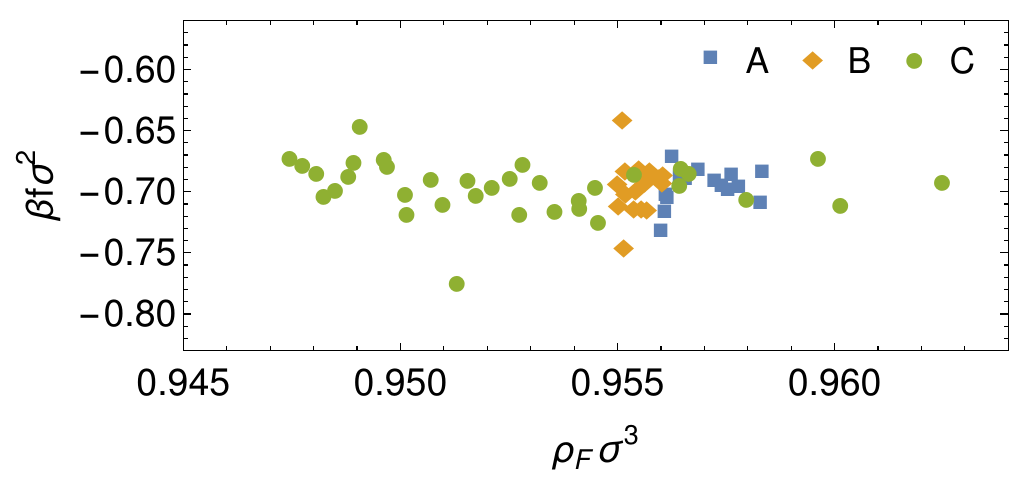} 
\end{tabular}
    \caption{\label{fig:sphericalsurfacestress} 
    \textbf{a)} Radial profile of the total pressure, normal pressure, and tangential pressure for one of the largest nuclei investigated ($\rho_F\sigma^3=0.94824$). Note that $P = ( P_\perp + 2P_\parallel)/3$. The dashed lines indicate the average pressures in the ``bulk'' crystal and fluid phases.
    \textbf{b)} Surface stress as a function of the density of the fluid phase for all investigated nuclei.
    }
\end{figure}

To find equilibrium coexistence conditions, we then look for the initial density at which the crystal phase is unstrained. This happens when the pressure in the $z$-direction coincides with that of a bulk unstrained crystal at the same density $\rho_\text{init}$ \cite{smallenburg2024simple}. For the coexistence satisfying this criterion, we obtain $f$ using Eq. \ref{eq:fflat}, making use of the fact that for both the (100) and the (111) plane $f_{xx} = f_{yy}$:
\begin{equation}
    f = \frac{1}{2} L_z \left(P_{zz} - \frac{P_{xx}+P_{yy}}{2}\right).\label{eq:surfacestressflat}
\end{equation}
Note that for a planar interface, the interfacial free energy and surface stress are independent of the choice of dividing surface.
We obtain  $f = -1.0(1)k_BT/\sigma^2$ for the (111) plane and $f = -0.24(4)k_BT/\sigma^2$ for the (100) plane. Given that the results are quite sensitive to the accurate determination of the equilibrium coexistence conditions, these values are in good agreement with the ones reported by Davidchack and Laird in Ref. \onlinecite{davidchack1998simulation}, i.e. $-0.71(13)k_BT/\sigma^2$ for (111) and $-0.17(6)k_BT/\sigma^2$ for (100).  The negative values of $f$ confirm that it is indeed reasonable to expect that the spherically averaged $f$ for a spherical nucleus is negative as well, explaining the sign of the pressure difference between the inside and outside of the nucleus in the previous section.

To obtain a more direct estimate, we also measure the spherically averaged surface stress using the simulations of spherical nuclei discussed in the previous sections. Specifically, we measure the radial profiles of the normal and tangential pressures for each nucleus. 
A typical example of these pressure profiles is shown in Fig. \ref{fig:sphericalsurfacestress}a. 
Then, using Eq. \ref{eq:fspherical} with the equimolar radius $R = R_e$ as our dividing surface, we calculate $f$ in for each nucleus. Note that $R_e$ has the advantage that it can be directly determined from the densities of the fluid phase, solid phase, and global system, which are all known quantities in each simulation. In Fig. \ref{fig:sphericalsurfacestress}b we plot the behavior of $f$ as a function of the density of the fluid phase. Within our error bars, the surface stress is approximately constant, around $\beta f \sigma^2 \simeq -0.7$. This is in good agreement with our estimate for $f^*$ based on the pressure difference in Section \ref{sec:pressure}. 
Note that the fluctuations in $P_\parallel$ shown in Fig. \ref{fig:sphericalsurfacestress}a complicate the accurate determination of $f$. Moreover, it should be kept in mind that our measurement of $f$ relies on the approximation that the interface is perfectly spherical and stationary during our simulations. Fluctuations in the nucleus shape will affect both the local curvature and the position of the interface, which might introduce systematic errors in the determination of $f$ for finite-sized clusters. Hence, the results in Fig. \ref{fig:sphericalsurfacestress}b should be considered an estimate rather than an exact determination.

Note that, naively, one could also have obtained an approximation for $f$ by treating the normal and tangential pressure profiles as if they belonged to a planar interface, ignoring the effects of curvature. In this planar approximation, $f$ could be obtained by integrating over the difference between the normal and tangential pressure profiles along the interface \cite{davidchack1998simulation}. 
This approximation results in values for $f$ that differ at most $\pm0.03 k_BT/\sigma^2$ with the values obtained via Eq. \ref{eq:fspherical} for the nuclei investigated.

%%%%%%%%%%%%%%%%%%%%%%%%%%%%%%%%%%%%%%%%%%%%%%%%%%%%%%%%%%%%%%%
%%%%%%%%%%%%%%%%%%%%%%%%%%%%%%%%%%%%%%%%%%%%%%%%%%%%%%%%%%%%%%%

\section{Chemical potential of the crystal phase and the interfacial free energy} \label{sec:mu}

The systems that we explore in this paper are in equilibrium.  Hence, the chemical potential must be the same (homogeneous) throughout the entire simulation box. The chemical potential of the fluid is easy to determine from its density profile: far away from the interface, the fluid must simply be a bulk fluid and its chemical potential can be determined from the equation of state.
For the crystal phase, an additional complication arises.  Because the crystal is a solid, deforming it at its boundaries inherently affects the lattice spacing deep inside the crystal. 
In comparison to the fluid, the lattice spacing provides an additional degree of freedom that can be tuned independently from the chemical potential. This was included in Eq. \ref{eq:omegatot} by the additional dependence of $\Omega_X$ on the unit cell size $v$ \footnote{Note that for simplicity, we only consider hydrostatic deformations of the lattice. This is consistent with the observation that our simulations show no sign of anisotropic compression of the crystal nucleus.}. The actual lattice spacing of the crystal nucleus is then set by a competition between the crystal phase and the interface. The crystal phase inside the nucleus favors a lattice spacing as close as possible to the bulk equilibrium value at chemical potential $\mu$. On the other hand, because the surface stress is negative, the interfacial free energy can be reduced by increasing the lattice spacing, favoring larger lattice spacings. This effect is particularly strong for small nuclei, where the surface-to-volume ratio of the nucleus is high.
As a result of this strain, the crystal phase we observe in our simulations does \textit{not} correspond to a bulk equilibrium crystal phase at the same density. Instead, it is a crystal under strain due to the presence of the spherical interface. This crystal necessarily has the same chemical potential as the surrounding fluid, but is stretched out by this strain, resulting in a lower density than a bulk crystal would have at the same chemical potential.

\begin{figure}[t!]
    \centering
    \includegraphics[width=\figwidthA]{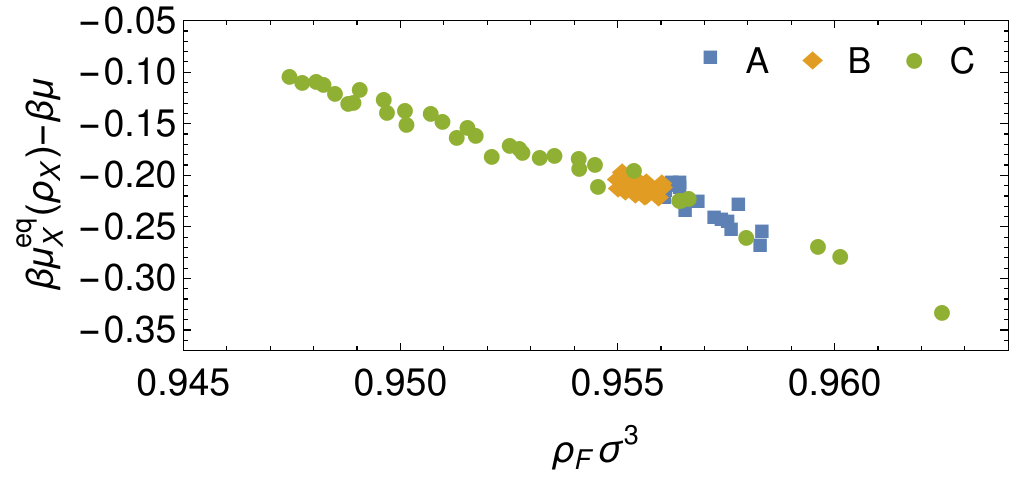}
    \caption{Chemical potential difference between a bulk equilibrium crystal at the same density as that of the crystal nucleus and the chemical potential of the fluid for all investigated nuclei.}
    \label{fig:mu}
\end{figure}

A natural question is then: how does this strained crystal differ from an equilibrium crystal at the same density? In Fig. \ref{fig:mu}, we plot the difference between the chemical potential of the fluid $\mu$ and that of an equilibrium bulk crystal with the same density as the one we measure inside the crystal nucleus ($\mu_X^\mathrm{eq}(\rho_X)$), as a function of the fluid density. We clearly find a negative apparent chemical potential difference, consistent with the idea that the crystal density is ``too low'' for its chemical potential. This also indicates that the crystal phase inside the nucleus must differ from an equilibrium crystal phase at the same density. 
Leaving aside the possibility of anisotropic deformations of shape of the unit cell, the most obvious possibility is that the existence of defects is responsible for this apparent paradox. 

The primary effect of changing the chemical potential of a crystal while keeping its lattice parameters fixed is a change in the concentration of point defects inside the crystal. In the case of monodisperse hard spheres, the dominant point defects are vacancies, which in an equilibrium near the melting point occur in a concentration of approximately $10^{-4}$ defects per lattice site \cite{bennett1971studies,pronk2001point,pronk2004large}. In the following, we explore how such vacancies affect the chemical potential of the crystal nucleus.

%%%%%%%%%%%%%%%%%%%%%%%%%%%%%%%%%%%%%%%%%%%%%%%%%%%%%%%%%%%%%%%

\begin{figure}[t!]
    \centering
    \includegraphics[width=\figwidthA]{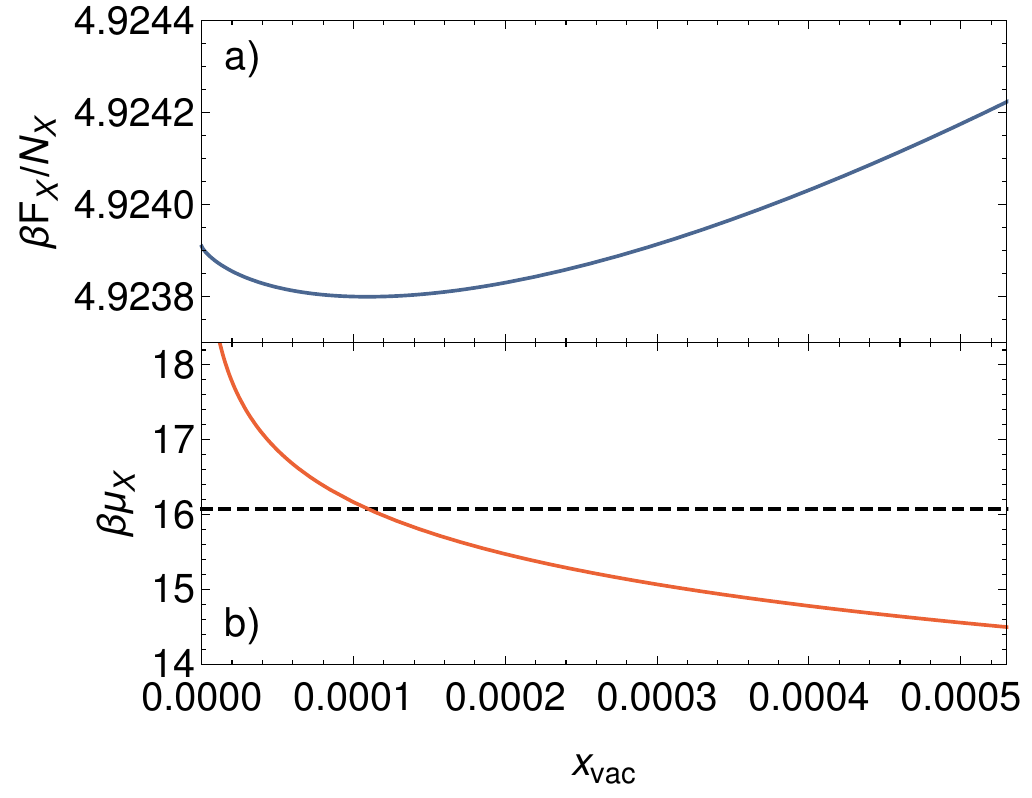}
    \caption{\textbf{a)} Helmholtz free energy of a hard-sphere crystal at the melting density as a function of the vacancy concentration. \textbf{b)} The chemical potential of the crystal at the melting density as a function of the vacancy concentration. The dashed line indicates the equilibrium chemical potential at this density.}
    \label{fig:defects}
\end{figure}

\subsection{Theory and Results}

Consider a crystal nucleus with fixed volume $V_X$ and number of lattice sites $M$, while the number of particles $N_X$ can be varied by exchanging particles with the surrounding fluid phase. The crystal nucleus can then tune its chemical potential by changing the defect concentration. 
Since for hard spheres, vacancies are much more frequent than interstitials, we only consider the possibility of vacancies.
Assuming non-interacting vacancies, which is reasonable at low vacancy concentrations, the Helmholtz free energy of the crystal nucleus is given by \cite{frenkelbook}
\begin{equation}\label{eq:totalFx}
\begin{split}
    F_X(M,V_X; N_X) =& F_X^\text{df}(M,V_X) + (M-N_X)f_\text{vac}\\
        &+ F_\text{conf}(M,N_X),
\end{split}
\end{equation}
where the first term, $F_X^\text{df}$, is the free energy of a defect-free crystal, the second term is the free-energy cost associated with creating $M-N_X$ vacancies at specific lattice sites, and the last term is the configurational free energy given by
\begin{equation}\label{eq:Fconfig}
\begin{split}
    F_\text{conf}(M,N_X) = M & \left[ x_\text{vac} \log x_\text{vac} \right. \\
        & \left. + (1-x_\text{vac}) \log(1-x_\text{vac})\right],
\end{split}
\end{equation}
with $x_\text{vac}=(M-N_X)/M$ the vacancy concentration. 
Using that $x_\text{vac}=1.10(2)\cdot10^{-4}$ is the equilibrium concentration of vacancies in a hard-sphere crystal at melting \cite{pronk2004large}, we obtain $f_\mathrm{vac} = -\log x_\text{vac} - \mu_X^\text{eq}(\rho_X) = -6.956 k_BT$. Combining this with the known free-energy behavior of a defect-free hard-sphere crystal, 
we plot in Fig. \ref{fig:defects}a the free energy of a crystal as a function of defect concentration at a fixed density equal to the melting density. Clearly, for the hard-sphere crystal the defects have a negligible effect on the free energy, hence \begin{equation}
\mathcal{F}_X(\rho_X, x_\mathrm{vac} = 0) \simeq \mathcal{F}_X(\rho_X, x_\mathrm{vac}^\mathrm{eq}),
\end{equation}
where $\mathcal{F}_X = F_X/N_X$, and $x_\mathrm{vac}^\mathrm{eq}$ is the equilibrium defect concentration. Moreover, as the pressure is the derivative of the free-energy with respect to the volume, it must be similarly unaffected by defects: $P_X(\rho_X, x_\mathrm{vac} = 0) \simeq P_X(\rho_X, x_\mathrm{vac}^\mathrm{eq})$.
In Section \ref{sec:pressure}, we indeed observed that the pressure inside the crystal nucleus agrees well with the pressure obtained from the bulk equation of state evaluated at the density of the crystal nucleus.

In contrast, the chemical potential of the crystal, which is given by 
\begin{equation}
    \mu_X = \left(\frac{\partial F_X}{\partial N_X}\right)_{M,V_X} = - f_\text{vac}
    - \log x_\text{vac} + \log(1-x_\text{vac}),
\end{equation}
is greatly affected by the presence of defects.
In Fig. \ref{fig:defects}b, we plot $\mu_X$ as a function of $x_\text{vac}$. 
If we compare this to the chemical potential differences observed in Fig. \ref{fig:mu} (which are on the order of 0.2$k_BT$), we see that only tiny changes in defect concentration are required to change the chemical potential of the crystal to match that of the fluid. In practice, such small variations of the already very low defect concentration would be essentially impossible to measure in our simulations. Hence, although a shift in chemical potential on the order of $0.2 k_B T$ might seem significant, in practice its effects can be readily accounted for by nearly imperceptible changes to the defect concentration.

When looking at Fig. \ref{fig:defects}b, it might seem puzzling at first glance that the chemical potential diverges in the limit of zero defects. This might appear to conflict with our usual treatment of free energies of defect-free crystals, where we typically calculate the ``defect-free'' chemical potential via the relation $\mu^{\mathrm{df}}_X = \mathcal{F}_X + P_X/\rho_X$. This is the chemical potential associated with a system where the only way particles can be added or removed is by simultaneously adding or removing a lattice site from the system. When vacancies are allowed, the system can also change the number of particles by creating or annihilating a defect. In this picture, putting a defect-free crystal in contact with a particle reservoir would indeed always lead to a flow of particles out of the system, reflecting the diverging chemical potential \footnote{Note that if interstitials are allowed as well, the true divergence is avoided, as there is a finite (but very high) chemical potential where vacancies and interstitials balance each other, resulting in a net zero flow between crystal and particle reservoir.}. Importantly, the chemical potential at the equilibrium vacancy concentration is essentially identical to $\mu^{\mathrm{df}}$, as
\begin{eqnarray}
\mu_X^\mathrm{eq} (\rho_X) &=& \mathcal{F}_X(\rho_X;x^\mathrm{eq}_\mathrm{vac}) + \frac{P_X(\rho_X;x^\mathrm{eq}_\mathrm{vac})}{\rho_X}\label{eq:mudf1}\\
&\simeq&  \mathcal{F}_X(\rho_X;x_\mathrm{vac}=0) + \frac{P_X(\rho_X;x_\mathrm{vac}=0)}{\rho_X} \\
&=& \mu_X^\mathrm{df}(\rho_X).
\end{eqnarray}
Hence, we are justified in using the ``defect-free'' chemical potential instead of the equilibrium one for the purpose of e.g. determining phase boundaries.

We can relate the results in Fig. \ref{fig:mu} on the behavior of $\mu_X^\mathrm{eq}(\rho_X) - \mu$ to the properties of the interface by using again the knowledge that the Helmholtz free energy and pressure of the crystal are only weakly affected by vacancies. Specifically, using Eq. \ref{eq:mudf1}, we can write
\begin{eqnarray}
\mu_X^\mathrm{eq}(\rho_X) - \mu 
&\simeq&
\mathcal{F}_X(\rho_X;x^\mathrm{eq}_\mathrm{vac}) + \frac{P_X(\rho_X;x^\mathrm{eq}_\mathrm{vac})}{\rho_X}-\mu \nonumber\\
&=& \frac{\Omega_X}{N_X} + \frac{P_X}{\rho_X} = \frac{\omega_X+ P_X}{\rho_X}\nonumber\\
&=& \frac{2v}{R}(f-\gamma),
\end{eqnarray}
where in the last step we have used that for small defect concentrations $v=1/\rho_X$, as well as Eqs. \ref{eq:min_v} and \ref{eq:f}. 
Given that we already know $f$ for our system, this provides us with a way of calculating $\gamma$. Specifically, 
\begin{equation}
    \gamma \simeq
    f - \frac{R}{2v}(\mu_X^\mathrm{eq}(\rho_X) - \mu ).
\end{equation}

\begin{figure}[t]
    \centering
    \includegraphics[width=\figwidthA]{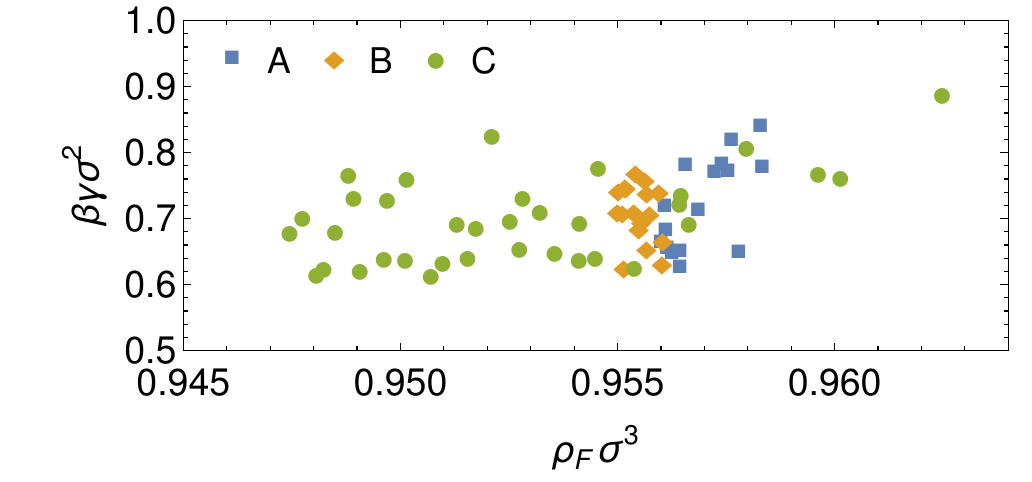}
    \caption{Interfacial free energy as a function of the density of the fluid phase for all investigated nuclei.}
    \label{fig:gamma}
\end{figure}

To see how $\gamma$ depends on the supersaturation of the system, in Fig. \ref{fig:gamma} we plot $\gamma$ as a function of the density of the fluid.
Despite the considerable scatter among data points, in general, we observe a very weak increase in $\gamma$ with increasing fluid density (and thus decreasing nucleus size). We note, however, that this determination of $\gamma$ relies on our earlier determination of $f$, which is likely to introduce some inaccuracy as discussed in Section \ref{sec:stress}.

%%%%%%%%%%%%%%%%%%%%%%%%%%%%%%%%%%%%%%%%%%%%%%%%%%%%%%%%%%%%%%%
%%%%%%%%%%%%%%%%%%%%%%%%%%%%%%%%%%%%%%%%%%%%%%%%%%%%%%%%%%%%%%%

\section{Thermodynamic model of the spherical crystal--fluid interface}\label{sec:model}

The wealth of data we have available here on large equilibrated  crystal nuclei allows us to take a closer look at the behavior of $\gamma$ both at coexistence and as a function of the supersaturation. In this section we attempt to use this wealth of data to determine $\gamma$ as a function of the chemical potential $\mu$,  the equimolar radius $R_e$ of the cluster and the lattice spacing $v$ in the crystal.

%%%%%%%%%%%%%%%%%%%%%%%%%%%%%%%%%%%%%%%%%%%%%%%%%%%%%%%%%%%%%%%

\subsection{Theory and Results}

As shown in Section \ref{sec:pressure} (Eqs. \ref{eq:omegatot} and \ref{eq:definegamma}), the thermodynamics of a system containing a critical nucleus are completely described by the thermodynamics of the bulk fluid ($\Omega_F(\mu,V_F)$), the bulk crystal ($\Omega_X(\mu,V_X,v)$), and the interface ($\gamma(\mu,v,R)$). In other words, if we have expressions for all three of these free energies, we can predict all thermodynamic properties of a system containing a critical nucleus. In particular, Eqs. \ref{eq:min_Vx} and \ref{eq:min_v} can be solved to obtain e.g. the equimolar radius and the crystal pressure, for any choice of the fluid chemical potential $\mu$.

For the hard-sphere system, we have excellent knowledge of the thermodynamics of the fluid via its equilibrium equation of state. As a result, we can readily evaluate $\Omega_F$ and its derivatives at any state point. Here, we use the KLM equation of state of Ref. \onlinecite{kolafa2004accurate} for the hard-sphere fluid.

For the crystal phase, the thermodynamics are more complex. Although we have excellent knowledge of the equation of state and free energy of a defect-free crystal \cite{}, we need to account for the effect of defects on $\Omega_X$ as well. We can address this by making the well-established assumption that the density, pressure and Helmholtz free energy $F$ are essentially unaffected by defects in the crystal (see Sec. \ref{sec:mu}).  In this approximation, we can write:
\begin{eqnarray}
    \rho_X &\simeq& 1/v ,\\
    P_X &\simeq& P_X^\mathrm{df}(\rho_X) ,\\
    \omega_X &=& \frac{\Omega_X}{V_X} \simeq \left({F_X^\mathrm{df}(\rho_X)} - \mu N\right) \frac{\rho_X}{N},
\end{eqnarray}
where the superscript $^\mathrm{df}$ refers to the properties of a bulk defect-free crystal. 
Here, we use Speedy's equation of state for the hard-sphere crystal \cite{speedy1998pressure} and use the excess Helmholtz free energy from Ref. \onlinecite{polson2000finite} as a reference point for obtaining $F_X^\mathrm{df}(\rho_X)$.

For the interface, we do not have a well-established functional form for $\gamma(\mu, v, R)$. However, the wealth of simulation data we have available on large equilibrated crystal nuclei allows us to fit an approximate function to $\gamma$. To this end, we make the following ansatz, based on a second-order Taylor expansion around the infinite-nucleus coexistence value $\gamma_0$:
\begin{eqnarray}
    \beta\gamma(\mu,v,R_e)\sigma^2 &=& \beta\gamma_0\sigma^2 + 
    c_v \frac{v-v_\mathrm{coex}}{\sigma^3} \nonumber\\
    \label{eq:ansatz}
    &&+ c_{vv} \left(\frac{v-v_\mathrm{coex}}{\sigma^3}\right)^2  +c_R\frac{\sigma}{R_e},
\end{eqnarray}
where $\gamma_0$ and the constants $c$ are unknown fit parameters. Note that the use of the equimolar surface ensures that the terms in the expansion scaling with $|\mu - \mu^{\mathrm{coex}}|$ must vanish to ensure
\begin{eqnarray}
    \left( \frac{\partial \gamma}{\partial \mu} \right)_{v,R} = -\frac{N_S}{A} = 0.
\end{eqnarray}

\begin{figure}[t!]
    \centering
    \includegraphics[width=\figwidthA]{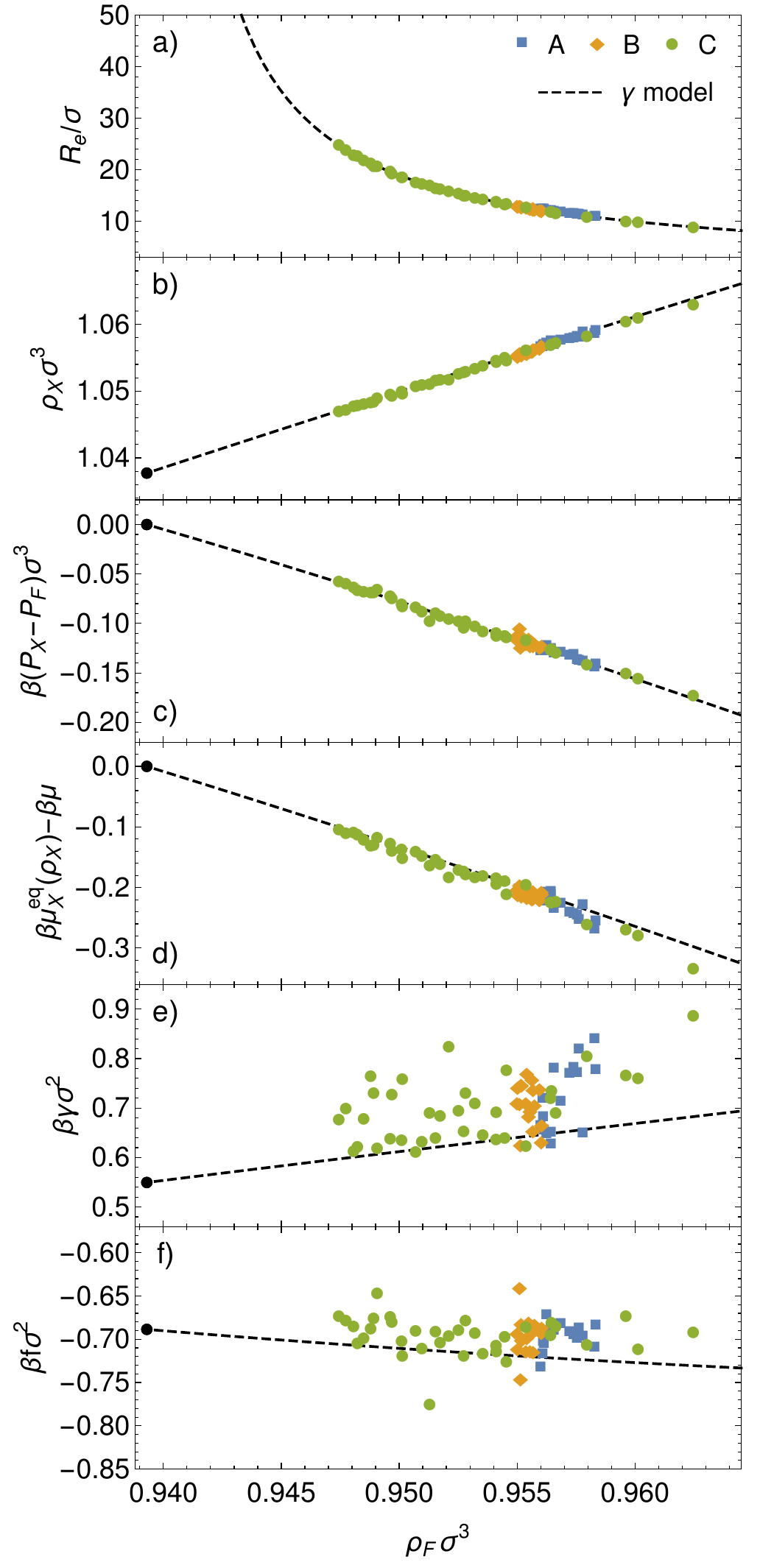}
    \caption{Different thermodynamic properties as a function of the density of the fluid for all investigated nuclei. In \textbf{a)} the equimolar radius, \textbf{b)} the density of the crystal nucleus, \textbf{c)} the pressure difference between the crystal nucleus and surrounding fluid, \textbf{d)} the chemical potential difference between a bulk equilibrium crystal at the same density as that of the crystal nucleus and the chemical potential of the fluid, \textbf{e)} the interfacial free energy, and \textbf{f)} the surface stress.
    The data points are the same results as in Figs. \ref{fig:bulkvalues}, \ref{fig:sphericalsurfacestress}b, \ref{fig:mu}, and \ref{fig:gamma}, but the figures now also include the result from the theoretical model with the fitted functional form of $\gamma$ (dashed lines). The black dots indicate the values at the freezing density, i.e. in the limit of an infinite nucleus. %$\rho_F=0.9393$
    }
    \label{fig:fitplots}
\end{figure}

Given a trial set of fit parameters, this ansatz allows us, for any fluid chemical potential $\mu$, to determine the equimolar radius $R_e$ and the pressure difference between the fluid and crystal $\Delta P$, using Eqs. \ref{eq:min_Vx} and \ref{eq:min_v}. We can compare these values to our measured equimolar radii and pressure differences from Section  \ref{sec:pressure} to optimize our trial fit parameters. To this end, we use a least-squares optimization, minimizing the relative squared prediction error in $R_e$ and $\Delta P$, summed over all investigated nuclei. 
Our resulting set of parameters results in the following fit: 
$\gamma_0 = 0.5496 k_BT / \sigma^2$, $c_v = -0.857$, $c_{vv}=3.078$, $c_R = 0.992$. 
Note that the value we obtain for $\gamma_0$ is in reasonable agreement with past estimates of the spherically averaged interfacial free energy \cite{davidchack2000direct, cacciuto2003solid, mu2005anisotropic, davidchack2010hard, hartel2012tension, sanchez2021fcc}, which range from $0.56 k_B T/\sigma^2$ to $0.66k_BT/\sigma^2$.

Based on the fitted functional form of $\gamma$, together with the thermodynamics of the bulk fluid and crystal, we can then predict all other thermodynamic aspects of the nucleus. In Fig. \ref{fig:fitplots}, we plot the equimolar radius, crystal density, pressure difference, interfacial free energy, and surface stress as a function of the fluid density, and compare the results to our simulation data. We find good agreement in all cases. 

Note that we have tried several functional forms for Eq. \ref{eq:ansatz}, by including higher-order terms in the Taylor expansion. However, these did not lead to large changes in our predictions. Hence, we here keep the lowest-order expansion that leads to a good fit of all of our simulation data. Note that since this is an expansion in the limit of large nucleus sizes, our expression for $\gamma$ is likely not accurate for significantly smaller nuclei (i.e. higher supersaturations) than the ones we used in our fitting procedure. Nonetheless, our expression should provide a convenient description of the interfacial thermodynamics of sufficiently large nuclei.

%%%%%%%%%%%%%%%%%%%%%%%%%%%%%%%%%%%%%%%%%%%%%%%%%%%%%%%%%%%%%%%
%%%%%%%%%%%%%%%%%%%%%%%%%%%%%%%%%%%%%%%%%%%%%%%%%%%%%%%%%%%%%%%

\section{Classical nucleation theory and free-energy barriers} \label{sec:cnt}

%%%%%%%%%%%%%%%%%%%%%%%%%%%%%%%%%%%%%%%%%%%%%%%%%%%%%%%%%%%%%%%

A crucial quantity in the study of nucleation processes is the height of the nucleation barrier, also known as the nucleation work. This quantity represents the free-energy cost of creating a critical nucleus out of the supersaturated fluid. In this section, we first show that the familiar expressions for the nucleation work from classical nucleation theory  still apply when taking into account strains on the crystal up to linear order \cite{mullins1984thermodynamic}. We then use simulations and the nucleation theorem \cite{kashchiev1982relation, oxtoby1994general, bowles2000molecular, richard2018crystallizationII} to calculate the nucleation work for critical nuclei at a range of supersaturations, and finally use this data to improve the thermodynamical model from Section \ref{sec:model}.

\subsection{Theory}
The work required to create a crystal nucleus is given by the difference between a system containing the nucleus and system of pure fluid.  In the grand-canonical ensemble, the nucleation work can be written as
\begin{eqnarray}
    \Delta\Omega &=& \Omega_\text{tot}(\mu,V;V_X,v) - \Omega_F(\mu, V) \label{eq:DeltaOmega}\\
    &=& \Omega_F(\mu,V_F) + \Omega_X(\mu,V_X,v) \nonumber\\ &&+ \gamma(\mu,R,v) A - \Omega_F(\mu, V) \nonumber\\
    &=& -P_F V_F + \omega_X V_X + \gamma A + P_F V\nonumber\\
    &=& \gamma A + (\omega_X + P_F)V_X. \label{eq:cntwork}
\end{eqnarray}
To connect to classical nucleation theory (CNT), we make the reasonable approximation that the crystal phase inside the nucleus is not strongly distorted with respect to its equilibrium lattice spacing $v^\mathrm{eq}(\mu)$. 
Within this approximation, $\omega_X$ is given by (minus) the pressure of the bulk crystal at the chemical potential of the fluid, i.e. $\omega_X(\mu,v)=-P_X^\mathrm{eq}(\mu)$, as shown in Eq. \ref{eq:Pthermodynamic}.
Using this, we can rewrite Eq. \ref{eq:cntwork} as one of the familiar CNT expressions
\begin{equation}
    \Delta\Omega \simeq \gamma A - \Delta P(\mu) V_X, \label{eq:CNT1}
\end{equation}
where $\Delta P(\mu)$ is the pressure difference between the two phases at equal chemical potential. 

If we additionally assume that the crystal density $\rho_X$ is approximately constant in the pressure regime containing $P_F$, $P_X$, and $P_X^\mathrm{eq}(\mu)$, we can write:
\begin{eqnarray}
    \mu_X(P_F) &\simeq& \mu_X(P_X^\mathrm{eq}(\mu)) - \left.\partdev{\mu_X}{P}\right|_{P_X} (P_X^\mathrm{eq}(\mu) - P_F)\nonumber\\
    &=& \mu - \rho_X (P_X^\mathrm{eq}(\mu) - P_F).
\end{eqnarray}
Hence, we can substitute $\Delta P$ in Eq. \ref{eq:CNT1} and obtain another familiar expression from CNT:
\begin{equation}
    \Delta\Omega \simeq \gamma A - \Delta \mu(P_F) N_X,
\end{equation}
where $\Delta \mu(P_F)$ is the chemical potential difference between the two phases at the fluid pressure.

If we now consider the derivative of the nucleation work with respect to $\mu$, then:
\begin{eqnarray}
\diffix{\Delta \Omega}{\mu}{V} &=&
\diffix{\Delta \Omega_\text{tot} }{\mu}{V,V_X,v} - \diffix{\Omega_F}{\mu}{V} \nonumber\\ 
&=& -N + \rho_F V \equiv -\Delta N, \label{eq:nuctheorem}
\end{eqnarray}
where in the first line we have used that the derivative of the grand potential with respect to $v$ and $V_X$ vanishes at the saddle point. The quantity $\Delta N = N - \rho_F V$ is the excess number of particles in the system with a nucleus, in comparison to a pure fluid system under the same conditions. Equation \ref{eq:nuctheorem} is sometimes called the nucleation theorem \cite{kashchiev1982relation, oxtoby1994general, bowles2000molecular, richard2018crystallizationII}. By integrating it, we can calculate the work required to create a critical nucleus:
\begin{equation} \label{eq:integration}
    \Delta \Omega(\mu) = \Delta \Omega(\mu^\mathrm{ref}) + \int_{\mu^\mathrm{ref}}^\mu \dd \mu^\prime \Delta N(\mu^\prime), 
\end{equation}
provided we know $\Delta N(\mu)$ as well as the nucleation work $\Delta \Omega(\mu^\mathrm{ref})$ at a reference chemical potential $\mu^\mathrm{ref}$.

Often with nucleation studies one works in the isobaric-isothermal (Gibbs) ensemble. In that case the Gibbs free-energy difference between the nucleating system and a pure metastable fluid is given by a Legendre transform of $\Delta \Omega$:
\begin{eqnarray}
    \Delta G &=& G_\text{tot}(N,P;N_X,M) - G_F(N,P)\\
    &=& \Omega_\text{tot}(\mu,V;V_X,v) + N \mu + P V \nonumber\\
    &&- \Omega_F(\mu,V) - N \mu - P V\\
    &=& \Delta \Omega.
\end{eqnarray}
Here, the pressure and chemical potential both correspond to those of the parent fluid phase. Hence, the nucleation work is the same in the grand-canonical and isobaric-isothermal ensembles.

%%%%%%%%%%%%%%%%%%%%%%%%%%%%%%%%%%%%%%%%%%%%%%%%%%%%%%%%%%%%%%%

\subsection{Methods}

One way of obtaining $\Delta \Omega$ (or equivalently $\Delta G$) is to directly use Eq. \ref{eq:cntwork}, using our knowledge of the thermodynamics of the two phases and the fitted $\gamma$ from Section \ref{sec:model} to evaluate it numerically. As an extra check, we can also obtain $\Delta \Omega$ via Eq. \ref{eq:integration}, where we take the reference point for the integration from past measurements of the nucleation work via umbrella sampling simulations \cite{filion2010crystal}. To do this, however, we require knowledge of $\Delta N(\mu)$ over a large range of chemical potentials, spanning from the relatively low supersaturations where umbrella sampling data is available ($\beta\mu \simeq 19.6$) to the regime where we performed our simulations of stable nuclei ($\beta \mu \lesssim 17.6$). Filling in the gap between these limits requires additional simulations at intermediate supersaturations, where keeping a finite nucleus stable for long periods of time is not feasible.

To address this issue, we perform umbrella simulations using a hybrid simulation approach: we perform short simulation trajectories in the canonical ensemble using our EDMD code, and either accept or reject the trajectory based on a biasing potential $U_\mathrm{bias}$, given by:
\begin{equation}
    U_\mathrm{bias} = \kappa \left(n - n_\mathrm{target}\right)^2,
\end{equation}
where $\kappa$ is a spring constant, $n$ is the size of the nucleus based on bond-orientational order parameters (see Section \ref{ssec:pressuremethods}), %(see e.g. Ref. \onlinecite{filion2010crystal}), 
and $n_\mathrm{target}$ is the target nucleus size. Each simulation is initialized containing a spherical nucleus of approximately the target size, and after equilibration we measure the average size $\langle n \rangle$ reached by the system, as well as the global pressure. Our goal is not to sample the entire nucleation barrier, but rather to find, for a given system size, the global density $\rho$ where a nucleus of size $n_\mathrm{target}$ is stable without biasing. Hence, for a series of system sizes and choices of $n_\mathrm{target}$, we perform a series of simulations with different global densities, and locate the global density where $\langle n \rangle = n_\mathrm{target}$. Under this condition, the biasing potential is not exerting any effective force on the system, indicating that the nucleus would be at a saddle point in the free-energy landscape without the biasing potential, and is therefore either a stable or critical nucleus. In other words, the system is either at the maximum or minimum of the top plot in Fig. \ref{fig:cluster}a.  Note that this approach is similar in spirit to the interface pinning approach for flat interfaces \cite{pedersen2013direct}. The global pressure at this density then corresponds to the pressure of a fluid which can coexist with a nucleus of size $n_\mathrm{target}$.

We consider system sizes between $N=2916$ and $32000$, with $n_\mathrm{target}$ for each system size corresponding to $\alpha N$, with $\alpha \in \{0.04, 0.06, 0.08, 0.15\}$. As a spring constant, we use $\beta \kappa = 10^5 / N$. For each system size and $n_\mathrm{target}$, we perform simulations for a range of densities $\rho$, and measure both $n_\mathrm{target}$ and $P$. After discarding simulations where the nucleus melted or percolated the simulation box, we fit $\langle n\rangle$ as a function of $\rho$, and find the density $\rho^\mathrm{unbiased}$ where it equals $n_\mathrm{target}$. The associated pressure $P^\mathrm{unbiased}$ at this density is obtained by fitting $P(\rho)$ and evaluating it at $\rho^\mathrm{unbiased}$. From $P^\mathrm{unbiased}$ and the fluid equation of state, we can then directly calculate $\Delta N$:
\begin{equation}
    \Delta N = N - \rho_F(P) V,
\end{equation}
where $\rho_F(P)$ is simply the inverse of the equilibrium fluid equation of state. Additionally, from the equilibrium fluid equation of state we also know the corresponding chemical potential $\mu$, this giving us a set of points tracing out the desired function $\Delta N(\mu)$.

%%%%%%%%%%%%%%%%%%%%%%%%%%%%%%%%%%%%%%%%%%%%%%%%%%%%%%%%%%%%%%%

\begin{figure}[t!]
\begin{tabular}{l}
     a) \\[-0.4cm]
     \includegraphics[width=\figwidthA]{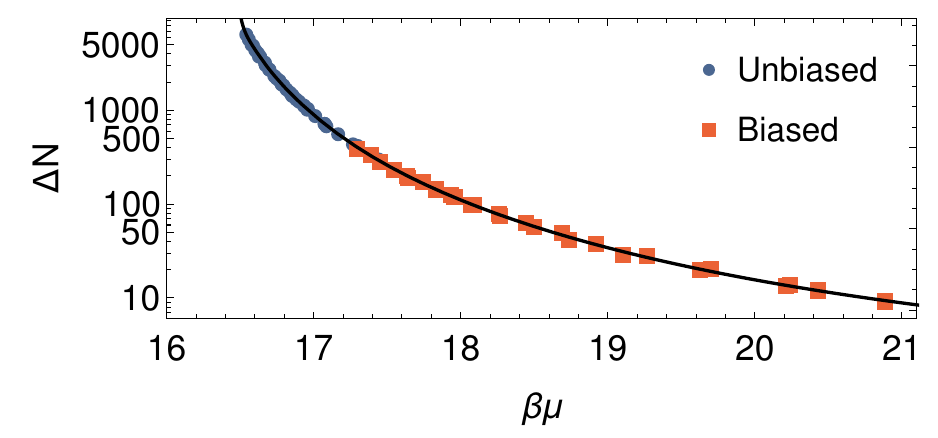} \\
     b) \\[-0.4cm]
    \includegraphics[width=\figwidthA]{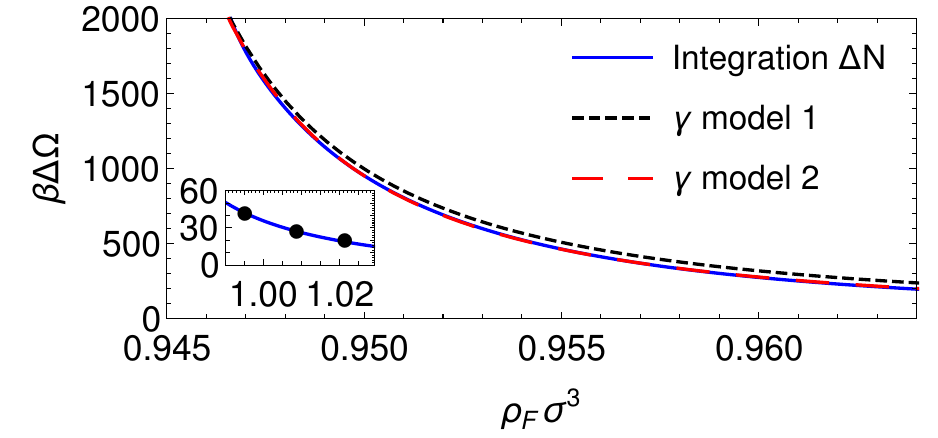} 
\end{tabular}
    \caption{\label{fig:DeltaN} 
    \textbf{a)} Excess number of particles as a function of the chemical potential. Blue dots are the results of the unbiased simulations (of nuclei C) of Section \ref{sec:pressure}. Red squares are data from the biased simulations. The black line is a fit (Eq. \ref{eq:deltaNfit}) to all data.
    \textbf{b)} Nucleation work as a function of the density of the fluid. The blue solid line indicates the result from Eq. \ref{eq:integration} and the dashed lines indicate the result from Eq. \ref{eq:cntwork} using the fitted functional form of $\gamma$. The black dashed line uses the original fitted $\gamma$ (as in Fig. \ref{fig:fitplots}), whereas the red longer-dashed line uses a $\gamma$ fitted with an additional loss term for $\Delta\Omega$.
    The inset shows the nucleation work (obtained via umbrella sampling) for hard spheres at higher supersaturations reported by Filion \textit{et al.} in Ref. \onlinecite{filion2010crystal} (black points).
    The blue line shows the results from Eq. \ref{eq:integration}, using the black point at $\rho_F\sigma^3=0.9952$ as reference point. 
    }
\end{figure}

\subsection{Results}
As a first step towards obtaining the nucleation work, we measure  $\Delta N(\mu)$ in both our biased simulations and from our previous nuclei. In Fig. \ref{fig:DeltaN}, we show the results for both simulations. All the data from different system sizes and target nucleus sizes collapse onto a single line, as expected. This allows us to fit $\Delta N$ using
\begin{equation} \label{eq:deltaNfit}
    \log \Delta N = \sum_{i=0}^8 \frac{c_i}{(\mu - \mu_\text{coex})^{i} },
\end{equation}
where the constants $c_i$ are the fit parameters. 
We then integrate the fit to obtain the nucleation work (Eq. \ref{eq:integration}), which is shown in Fig. \ref{fig:DeltaN}b as a blue solid line. The resulting nucleation barriers match closely those predicted for almost-hard spheres modeled via the Weeks-Chandler-Andersen potential \cite{richard2018crystallizationII}. 
It is also possible to predict the nucleation work from the functional model for $\gamma$ predicted in Section \ref{sec:model}, and Eq. \ref{eq:cntwork}.  The result is also shown in Fig. \ref{fig:DeltaN}b as the black dashed line. While the prediction shows the correct trend, the nucleation work predicted via this fit is approximately 40$k_B T$ off from the one obtained via integration. 

Given the excellent agreement with the thermodynamic parameters in Fig. \ref{fig:fitplots}, it is natural to wonder if the data we had contained sufficient information to fully determine not only $\gamma$, but also its functional dependence on $R$ and $v$. To test this, we refit all of our data, but now also including the new information on $\Delta \Omega$.  Specifically, we used a least squares optimization minimizing the relatively squared prediction error in $R_e$, $\Delta P$, and $\Delta \Omega$. The resulting $\gamma$ parameters are 
$\gamma_0 = 0.5645 k_BT / \sigma^2$, $c_v = -0.880$, $c_{vv}=2.190$, and $c_R = 0.511$. The slightly higher value of $\gamma_0$ is again consistent with past measurements \cite{davidchack2000direct, cacciuto2003solid, mu2005anisotropic, davidchack2010hard, hartel2012tension, sanchez2021fcc} of this quantity ($0.56 k_B T/\sigma^2$ to $0.66 k_B T/\sigma^2$).
The new fit is also shown in Fig. \ref{fig:DeltaN}b as the red longer-dashed line.  Clearly this new fit is able to capture the behavior of $\Delta \Omega$ excellently.  Interestingly, as we show in the Supplementary Material, the fit comes at no noticeable cost when it comes to fitting $R_e$ and $\Delta P$, indicating that our previous fit was indeed underdetermined.

%%%%%%%%%%%%%%%%%%%%%%%%%%%%%%%%%%%%%%%%%%%%%%%%%%%%%%%%%%%%%%%
%%%%%%%%%%%%%%%%%%%%%%%%%%%%%%%%%%%%%%%%%%%%%%%%%%%%%%%%%%%%%%%

\section{Conclusions}

In conclusion, we have extensively explored the thermodynamics of hard-sphere spherical crystal nuclei, both from a theoretical and simulation perspective. 
We examined the cause of the observed negative pressure difference between the inside and outside of the crystal nucleus, predicted the surface stress and interfacial free energy for spherical nuclei as a function of radius, examined the role of defects and chemical potential in the thermodynamics of the nuclei, and presented a simple thermodynamic model to capture the properties of the nucleus. We hope that our extensive study of hard-sphere critical nuclei will act as a foundation for future explorations into nucleation.

%%%%%%%%%%%%%%%%%%%%%%%%%%%%%%%%%%%%%%%%%%%%%%%%%%%%%%%%%%%%%%%
%%%%%%%%%%%%%%%%%%%%%%%%%%%%%%%%%%%%%%%%%%%%%%%%%%%%%%%%%%%%%%%

\section{Supplementary Material}
In the Supplementary Material we provide the derivation of Sec. \ref{sec:pressure} in the canonical ensemble, show how one can obtain the normal and tangential pressure profiles from the profile of total pressure, give some results on nuclei of HCP crystal, and compare the two fitted models for the interfacial free energy $\gamma$.

%%%%%%%%%%%%%%%%%%%%%%%%%%%%%%%%%%%%%%%%%%%%%%%%%%%%%%%%%%%%%%%
%%%%%%%%%%%%%%%%%%%%%%%%%%%%%%%%%%%%%%%%%%%%%%%%%%%%%%%%%%%%%%%

\section{Acknowledgements}
We would like to thank Rinske Alkemade  and Willem Gispen for useful discussions. L.F. and M.d.J. acknowledge funding from the Vidi research program with project number VI.VIDI.192.102 which is financed by the Dutch Research Council (NWO). 
C.V. acknowledges funding from Grant PID2022-136919NB-C31 of the Ministry of Science, Innovation and Universities (MICINN).
F.S. acknowledges funding from the Agence Nationale de la Recherche (ANR), grant ANR-21-CE30-0051.

\section{Data Availability Statement}
An open data package containing the (analyzed) data and other means to reproduce the results of the simulations will be published on Zenodo.

\bibliography{paper}% Produces the bibliography via BibTeX.

\end{document}

\preprint{APS/123-QED}

\title{Supplementary Material for ``Statistical mechanics of crystal nuclei of hard spheres''}% Force line breaks with \\

\author{Marjolein de Jager}
%\email{m.e.dejager@uu.nl}
\affiliation{Soft Condensed Matter and Biophysics, Debye Institute for Nanomaterials Science, Utrecht University, Utrecht, Netherlands}
\author{Carlos Vega}
\affiliation{Departamento de Qu\'imica F\'isica, Facultad de Ciencias Qu\'imicas, Universidad Complutense de Madrid, 28040 Madrid, Spain}
\author{Pablo Montero de Hijes}
\affiliation{Departamento de Qu\'imica F\'isica, Facultad de Ciencias Qu\'imicas, Universidad Complutense de Madrid, 28040 Madrid, Spain}
\author{Frank Smallenburg}
\affiliation{Universit\'e Paris-Saclay, CNRS, Laboratoire de Physique des Solides, 91405 Orsay, France}
\author{Laura Filion}
\affiliation{Soft Condensed Matter and Biophysics, Debye Institute for Nanomaterials Science, Utrecht University, Utrecht, Netherlands}%Lines break automatically or can be forced with \\

% \collaboration{CLEO Collaboration}%\noaffiliation

\maketitle

\onecolumngrid  %for single column paper

%\tableofcontents

\newcommand{\comment}[1]{{\color{red}{\bf #1}}}

\renewcommand{\thetable}{S\arabic{table}}
\renewcommand{\thefigure}{S\arabic{figure}}
\renewcommand{\theequation}{S\arabic{equation}}

\newcommand{\figwidthA}{0.48\linewidth}

%%%%%%%%%%%%%%%%%%%%%%%%%%%%%%%%%%%%%%%%%%%%%%%%%%%%%%%%%%%%%%%
%%%%%%%%%%%%%%%%%%%%%%%%%%%%%%%%%%%%%%%%%%%%%%%%%%%%%%%%%%%%%%%

This Supplementary Material includes the derivation of Section II of the main paper in the canonical ensemble, shows how one can obtain the normal and tangential pressure profiles from the profile of total pressure, gives some results on nuclei of HCP crystal, and compares the two fitted models for the interfacial free energy $\gamma$.

\section{Canonical derivation}

In Section II of the main paper, we revisit the theory of crystal-fluid coexistence using the grand-canonical ensemble, and derive the relation between the pressure inside a crystal nucleus and the pressure of the surrounding fluid (Eq. 8). Here, we derive the same relation, now using the canonical ensemble.
Using a Legendre transformation, we can write the total Helmholtz free energy of the system as
\begin{align}
    F_\text{tot}(N,V,T; N_X,V_X,v) &= \Omega_\text{tot}(\mu,V,T; V_X,v) + \mu N(\mu,V,T; V_X,v) , \\
    &= F_F(N_F,V_F,T) + F_X(N_X,V_X,T;v) + A \gamma(\mu,R,T,v) + \mu(N_F,V_F,T) A \Gamma(\mu,R,T,v) ,
\end{align}
where we used that $N=N_F+N_X+N_S$ with the number of excess interfacial particles given by $N_S=A\Gamma(\mu,R,T,v)$. Here, $\Gamma$ is the adsorption per unit area. 
The Helmholtz free energy has three internal variables: the number of particles in the nucleus $N_X$, the volume of the nucleus $V_X$, and the volume of a crystal unit cell $v$. The latter variable accounts for (point) defects in the crystal. 

In equilibrium, the free-energy landscape is locally flat with respect to the three internal degrees of freedom. Specifically,
\begin{align}
    \left(\frac{\partial F_\text{tot}}{\partial N_X}\right)_{N,V,V_X,v} &= 0 , \label{eq:Fconstraint1} \\
    \left(\frac{\partial F_\text{tot}}{\partial V_X}\right)_{N,V,N_X,v} &= 0 ,
    \label{eq:Fconstraint2} \\
    \left(\frac{\partial F_\text{tot}}{\partial v}\right)_{N,V,N_X,V_X} &= 0 .
    \label{eq:Fconstraint3}
\end{align}
Here and in all following equations, we omit the temperature $T$ dependence, as we do not consider any variations in temperature throughout this work.
Using that $\Gamma = -\left(\frac{\partial \gamma}{\partial \mu}\right)_{R,v}$, the first constraint results in the relation $\mu_X=\mu$, and the two others result in
\begin{align}
      P_F - P_X + \frac{2\gamma}{R} + \frac{3v}{R} \left(\frac{\partial \gamma}{\partial v}\right)_{\mu,R} + \left(\frac{\partial \gamma}{\partial R}\right)_{\mu,v} &= 0 , \\
      \omega_X +P_X - \frac{3v}{R} \left(\frac{\partial \gamma}{\partial v}\right)_{\mu,R} &= 0 ,
\end{align}
which are equivalent to Eqs. 6 and 7 of the main paper.

%%%%%%%%%%%%%%%%%%%%%%%%%%%%%%%%%%%%%%%%%%%%%%%%%%%%%%%%%%%%%%%
%%%%%%%%%%%%%%%%%%%%%%%%%%%%%%%%%%%%%%%%%%%%%%%%%%%%%%%%%%%%%%%

\section{Normal and tangential pressure profiles}

In the main paper, we explain how we directly measure the radial profiles of the normal and tangential pressures during the simulations. However, one can also obtain these profiles from the profile of the total pressure \cite{nakamura2011novel}. 
The total pressure is given by
\begin{equation} \label{eq:ptotal}
    P(r) = \frac{1}{3} \left( P_\perp(r) + 2P_\parallel(r) \right),
\end{equation}
where $P(r)$, $P_\perp(r)$, and $P_\parallel(r)$ indicate the total, normal, and tangential pressure profiles, respectively. 
Then, using that, for a spherical nucleus, mechanical equilibrium requires $\nabla\cdot\mathbf{P}=0$, i.e. \cite{rowlinson2013molecular}
\begin{equation}
    P_\parallel(r) - P_\perp(r) = \frac{r}{2} \frac{dP_\perp(r)}{dr},
\end{equation}
we can derive
\begin{equation} \label{eq:pnormal}
    P_\perp(r) = P_X + \frac{3}{r^3} \int_0^r dr' r'\,^2 \left( P(r') - P_X \right).
\end{equation}
Here, we used that deep inside the crystal nucleus the pressures are equal to the pressure of the bulk crystal, i.e. $P_\perp(0)=P(0)=P_X$.
Using the normal pressure profile of Eq. \ref{eq:pnormal}, the tangential pressure profile can then easily be obtained via Eq. \ref{eq:ptotal}, i.e. $P_\parallel(r) = \left( 3P(r) - P_\perp(r) \right)/2$.

\begin{figure}[h!]
     \centering
     \includegraphics[width=\figwidthA]{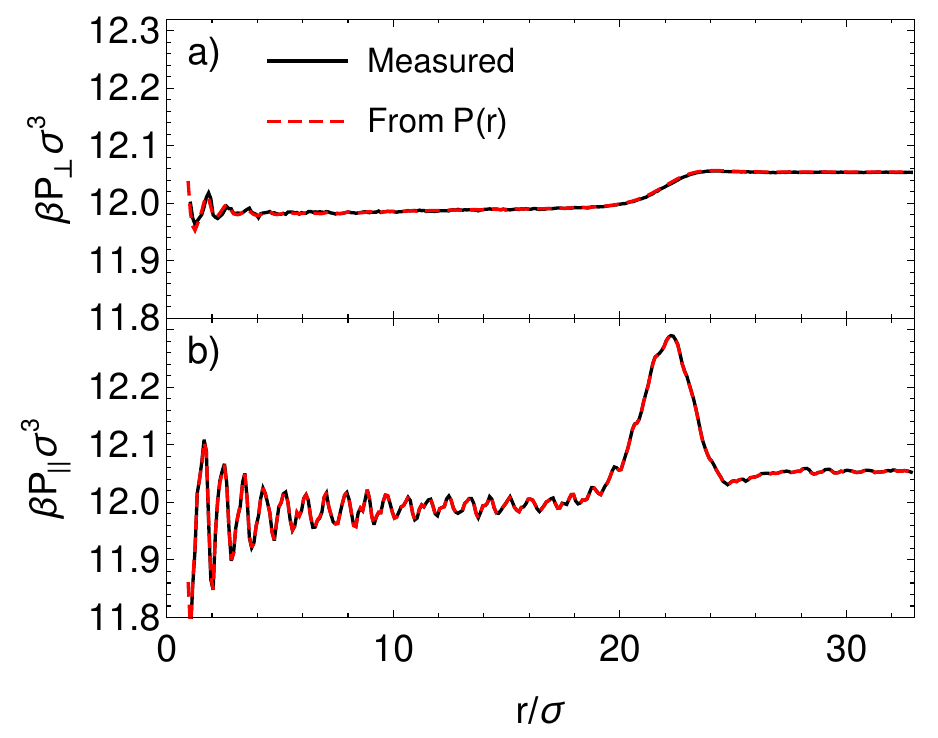}
     \caption{\label{fig:pprofiles}
     The \textbf{a)} normal and \textbf{b)} tangential pressure profiles for the same nucleus as in Fig. 5a of the main paper. The black solid lines indicate the profiles measured directly during the simulation and the red dashed lines indicate the profiles obtained from the total pressure profile (using Eq. \ref{eq:pnormal}).
     }
 \end{figure}

To demonstrate the effectiveness of this method, in Fig. \ref{fig:pprofiles}, we show the normal and tangential pressure profiles obtained using Eq. \ref{eq:pnormal} and compare them to the profiles measured directly during the simulation. We see that both profiles show excellent agreement between the two different methods.

%%%%%%%%%%%%%%%%%%%%%%%%%%%%%%%%%%%%%%%%%%%%%%%%%%%%%%%%%%%%%%%
%%%%%%%%%%%%%%%%%%%%%%%%%%%%%%%%%%%%%%%%%%%%%%%%%%%%%%%%%%%%%%%

\section{Nuclei of HCP crystal}
As the face-centered cubic (FCC) crystal is the stable crystal phase for hard spheres, all crystal nuclei considered in the main paper are nuclei of FCC crystal. However, for hard spheres, the hexagonal close-packed (HCP) crystal is metastable with only a very small free-energy difference (approximately $0.001 k_B T$ per particle near melting) with respect to the FCC crystal \cite{frenkel1984new, bolhuis1997entropy, mau1999stacking}.
Consequently, a (metastable) coexistence between a nucleus of HCP crystal and a fluid can be achieved.
To obtain systems containing HCP nuclei, we prepare 6 seeds of HCP crystal with sizes ranging from $8\cdot10^3$ to $34\cdot10^3$ particles and insert them in an equilibrated fluid. After a quick equilibration, we run the EDMD simulations for another $10^4\tau$ to obtain the radial density and pressure profiles. From these profiles we compute the thermodynamic properties of the systems (as explained in Sections II and III of the main paper). Note that, in order to obtain the chemical potential of the crystal $\mu_X^\text{eq}(\rho_X)$ (and consequently the interfacial free energy $\gamma$), we use the equation of state of the FCC crystal\cite{speedy1998pressure}.
In Fig. \ref{fig:hcpnewfit}, we show the resulting properties of the HCP nuclei (red triangles) and compare them to the ones obtained for the FCC nuclei (other markers). We find that all properties of the HCP nuclei agree with those of the FCC nuclei within the error margins of the measurement method.

%%%%%%%%%%%%%%%%%%%%%%%%%%%%%%%%%%%%%%%%%%%%%%%%%%%%%%%%%%%%%%%
%%%%%%%%%%%%%%%%%%%%%%%%%%%%%%%%%%%%%%%%%%%%%%%%%%%%%%%%%%%%%%%

\section{Comparing fitted models for $\gamma$}
\begin{figure}[h!]
\begin{tabular}{ll}
     \includegraphics[width=\figwidthA]{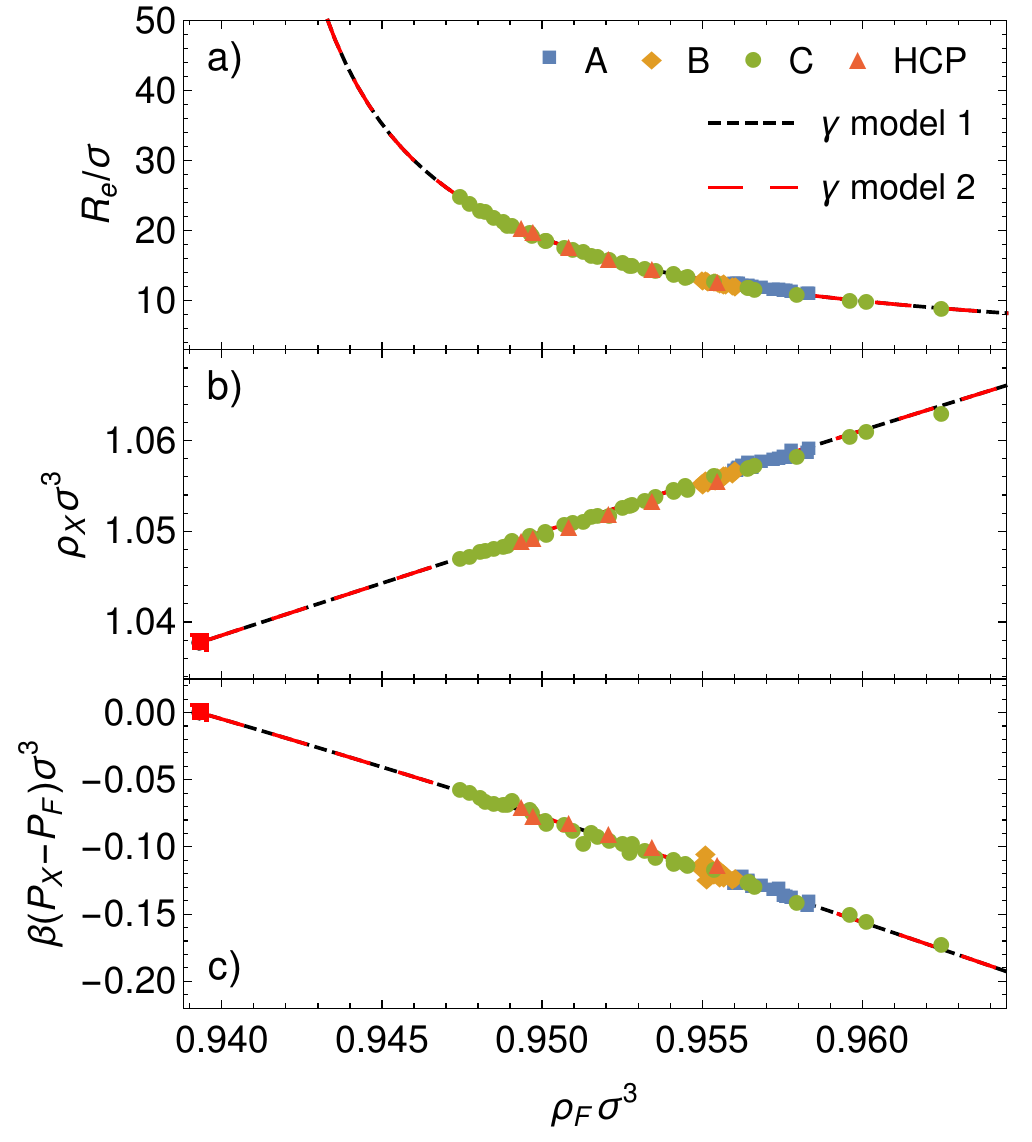} & \includegraphics[width=\figwidthA]{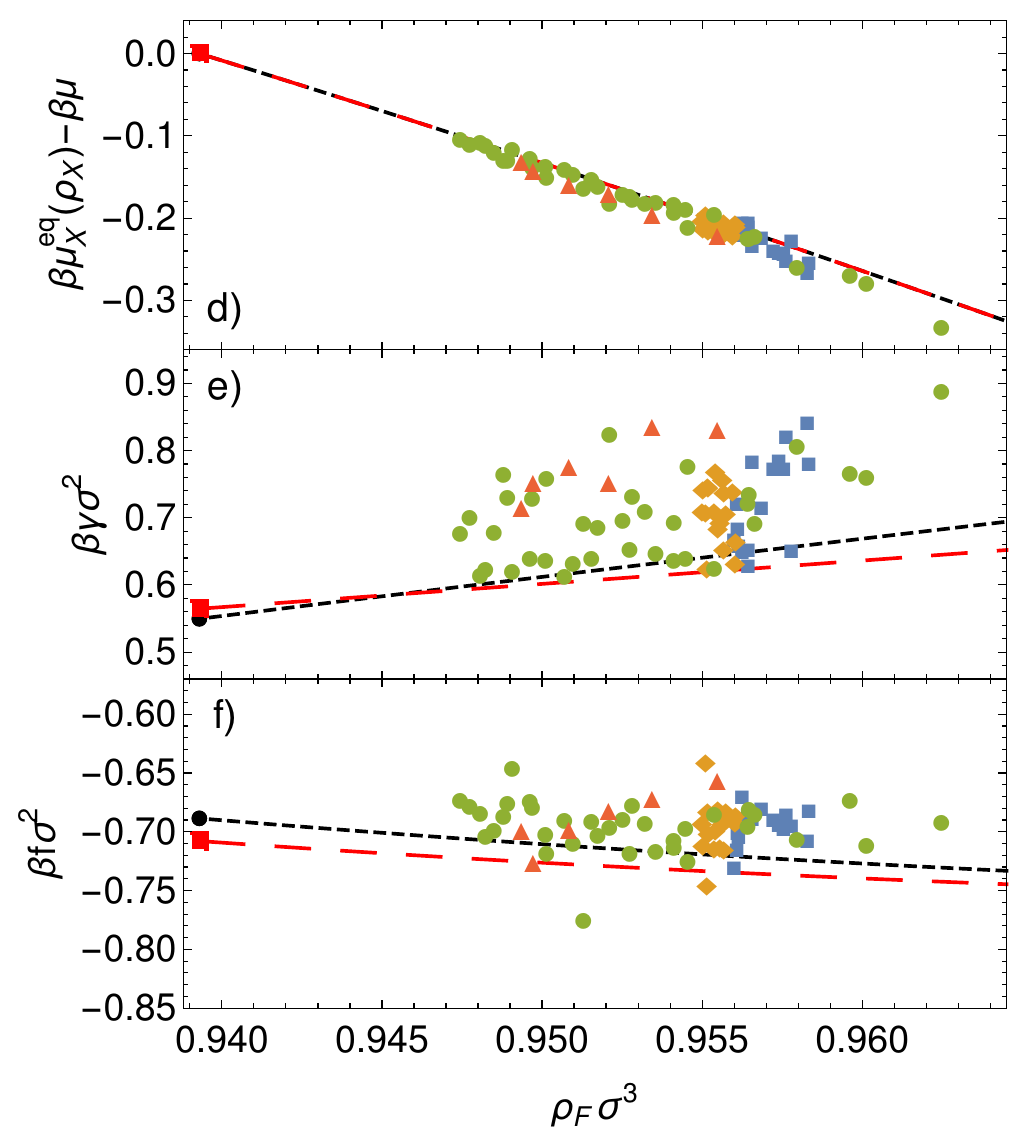} 
\end{tabular}
    \caption[width=1\linewidth]{\label{fig:hcpnewfit} 
    Different thermodynamic properties as a function of the fluid density for all investigated nuclei. These are the same figures as in Fig. 9 of the main paper. In addition, the figures contain the results of the HCP nuclei (red triangles), and the comparison between the original fitted $\gamma$ (black dashed line) and the refitted $\gamma$ (red longer-dashed line).
    In \textbf{a)} the equimolar radius, \textbf{b)} the density of the crystal nucleus, \textbf{c)} the pressure difference between the crystal nucleus and surrounding fluid, \textbf{d)} the chemical potential difference between a bulk equilibrium crystal at the same density as that of the crystal nucleus and the chemical potential of the fluid, \textbf{e)} the interfacial free energy, and \textbf{f)} the surface stress.
    }
\end{figure}

In Section VI.C of the main paper, we explain that we refit the functional form of the interfacial free energy $\gamma$ to better capture the behavior of the nucleation work $\Delta \Omega$.
In contrast to the fitted $\gamma$ of Section V, which is fitted using a least squares optimization minimizing the relative squared prediction error in the equimolar radius $R_e$ and the pressure difference $\Delta P$, the refitted $\gamma$ minimizes the relative squared prediction error in $R_e$, $\Delta P$ and $\Delta\Omega$. 
We show that the new fit captures the behavior of $\Delta\Omega$ excellently (Fig. 10b of the main paper), and claim that fit comes at no noticeable cost to fitting $R_e$ and $\Delta P$. 
Here, we show the results supporting this claim. In Fig. \ref{fig:hcpnewfit}, we show the thermodynamic properties of the system (as in Fig. 9 of the main paper) and compare the original fitted $\gamma$ (black dashed line) to the refitted $\gamma$ (red long-dashed line). We observe no noticeable difference in the models when it comes to $R_e$, $\Delta P$, the density of the crystal, or the chemical potential difference.

%%%%%%%%%%%%%%%%%%%%%%%%%%%%%%%%%%%%%%%%%%%%%%%%%%%%%%%%%%%%%%%
%%%%%%%%%%%%%%%%%%%%%%%%%%%%%%%%%%%%%%%%%%%%%%%%%%%%%%%%%%%%%%%

\bibliography{si_paper}% Produces the bibliography via BibTeX.